\documentclass[a4paper,fleqn]{cas-sc}

\usepackage[numbers]{natbib}

\usepackage{graphicx}
\usepackage{subcaption}
\usepackage[utf8]{inputenc}
\usepackage[export]{adjustbox}
\usepackage{wrapfig}
\usepackage{blindtext}

\usepackage{amssymb}
\usepackage{gensymb}

\usepackage{subcaption}

\usepackage{afterpage}

\usepackage{framed} 
\usepackage{multicol} 
\usepackage{nomencl} 
\makenomenclature

\renewcommand*\nompreamble{\begin{multicols}{2}}
\renewcommand*\nompostamble{\end{multicols}}

\usepackage{etoolbox}
\renewcommand\nomgroup[1]{%
  \item[\bfseries
  \ifstrequal{#1}{A}{Alphabetic}{%
  \ifstrequal{#1}{S}{Subscripts}{%
  \ifstrequal{#1}{O}{Other Symbols}{%
  \ifstrequal{#1}{B}{Abbreviation}{%
  \ifstrequal{#1}{N}{Numeration}{}}}}}%
]}

\def\tsc#1{\csdef{#1}{\textsc{\lowercase{#1}}\xspace}}
\tsc{WGM}
\tsc{QE}
\tsc{EP}
\tsc{PMS}
\tsc{BEC}
\tsc{DE}

\begin{document}
\let\WriteBookmarks\relax
\def\floatpagepagefraction{1}
\def\textpagefraction{.001}
\shorttitle{PVDF/paraffin fibers}
\shortauthors{A. Nikulin et~al.}

\title [mode = title]
{A facile approach for phase change material encapsulation into polymeric flexible fibers using microfluidic principles}                      


\author[1,2]{Mikel Duran}
\author[1]{Artem Nikulin\corref{cor1}}[orcid=0000-0002-3304-0506]\cormark[1]
\author[1]{Jean-Luc Dauvergne}
\author[1]{Angel Serrano}
\author[1,3]{Yaroslav Grosu}
\author[2]{Jalel Labidi}
\author[1,4]{Elena Palomo del Barrio}

\address[1]{Centre for Cooperative Research on Alternative Energies (CIC energiGUNE), Basque Research and Technology Alliance (BRTA), Alava Technology Park, Albert Einstein 48, 01510 Vitoria-Gasteiz, Spain}
\address[2]{University of the Basque Country (UPV/EHU), Plaza Europa 1, 20018 Donostia-San Sebastián, Gipuzkoa, Spain}
\address[3]{Institute of Chemistry, University of Silesia in Katowice, Szkolna 9 street, 40-006 Katowice, Poland}
\address[4]{Ikerbasque --- Basque Foundation for Science,  Plaza Euskadi 5, 48009 Bilbao, Spain}

\nonumnote{*Corresponding author: anikulin@cicenergigune.com}

\begin{abstract}
It is widely agreed that phase change materials (PCMs) are of high interest for sustainable energy future. Many of the applications require anti-leakage properties of PCM, that can be acomplished through PCM encapsulation. In this study, scalable and considerably simplified approach based on the microfluidics principles was successfully designed for polyvinylidene fluoride (PVDF) hollow- and for leakage-free paraffin-core/PVDF-sheath fibers production. The required device can be as simple as syringe+tube+glass capillary. The fibers were created by PVDF/N,N-Dimethylformamide (DMF) solution and PVDF/DMF/paraffin emulsion injection in water followed by solvent extraction process.

The proposed approach results in a hollow PVDF or PVDF/paraffin composite fibers with the PCM content between 32-47.5\% according to DSC and TGA measurements. SEM study of the fibers morphology has shown that PCM is in the form of slugs along the fibers. Such PCM distribution is maintained until the first melting cycle. Later, molten PCM spreads within the fiber under capillary forces that was captured by infrared camera. Elastic modules and stress vs. strain were measured to characterise mechanical properties of designed fibers. Finally, the composite fibers have shown outstanding retention capacity with only 3.5\% of PCM mass loose after 1000 melting/crystallisation cycles.

\end{abstract}


  


\maketitle

\section{Introduction}
\par
Sustainable energy production requires enhancement of potential materials for energy storage applications \cite{elvin2013advances} to promote wider spreading of renewable energy resources usage. Currently it is hard to underestimate the importance of phase change materials (PCMs) as an effective solution for a wide range of applications \cite{mehling2008heat}. Two mainstream areas of PCM implementation are thermal energy storage (TES) \cite{rathore2019potential,serrano2021tailored,prieto2019thermal,de2017lca,allouche2017dynamic} and thermal control \cite{grosu2020hierarchical,xie2018thermal,chen2020air}. Compared to other phase change materials, the attractiveness of solid-liquid PCMs is related to the large amount of heat stored and released during the solidification and melting steps, as well as the reduced temperature variation during those processes \cite{mehling2008heat}. Nonetheless, a serious engineering obstacle for solid-liquid PCM application is the possibility of leakages that can be driven by gravity, inertial and capillary forces. A variety of methods have been proposed in order to cope with that issue. The main idea is to encapsulate the PCMs or to stabilise their shape by confining them into a closed shell or into a porous matrix that prevents leaks \cite{cardenas2020systematic}. 

According to the size, capsules can be classified as macrocapsules (more than 1 mm), microcapsules (1$\cdot$10$^{-3}$---1 mm) and nanocapsules (less than 1$\cdot$10$^{-6}$ mm) \cite{jacob2015review}. As for the shape, most of the explored methods for PCMs encapsulation result in spherical or irregular shape capsules \cite{cardenas2020systematic}. However, physico-chemical and physical methods of PCMs encapsulation can be adopted for other shape outcome, i.e. foams or fibers \cite{cardenas2020systematic}. Recently, the methods for encapsulation of PCMs into fibers are attracting more attention of researchers due to their high applicability for textiles with thermal management \cite{shin2005development,yan2021flexible,chen2008novel,nguyen2011fabrication,xia2021nano}, thermal storage and protection \cite{hu2012encapsulation,golestaneh2016fabrication,babapoor2017coaxial,darzi2019thermal,xiang2019construction}. Moreover, an important advantage of fibers is that they can further be shaped using standard and well developed processes (weaving and knitting etc.) that promote their sooner production at industrial scale. The methods, key findings and typical issues of PCM encapsulation into fibers are briefly discussed below.

One of the earliest strategies explored in this field was the incorporation of encapsulated PCMs into traditional fibers or fabrics already available on the market. In this sense, Shin et al. \cite{shin2005development} prepared microcapsules based on an eicosane core in a melamine/formaldehyde shell by polymerization method. The microcapsules were later binded by polyurethane to polyester knit using pad-dry-cure (PDC) method. Authors reported from 0.91 to 4.44 J/g of latent heat gained by polyester knit depending on the load percentage of microcapsules. The main issue of such approach was the low resistance of the composite fabrics to wearing conditions. About 40 \% of latent heat remained after five launderings of composite fabric \cite{shin2005development}. In the study of Khoddami et al. \cite{khoddami2011improvement} a PDC method was applied for polyethylene glycol (PEG) incorporation into poly(lactic acid) and polyethylene terephthalate (PET) fabrics. Much higher latent heat of 43 J/g was achieved right after fabrics preparation. Nevertheless, strong reduction of latent heat was observed due to rinsing, washing and abrading. Wearing properties were significantly improved via hydrophilic coating by addition of dimethyloldihydroxyethyleneurea in PEG.

Another approach for PCM encapsulation into fibers is based on the coaxial jet elongation under electric field known as co-electrospinning \cite{sun2003compound,reneker2008electrospinning}. This technique was applied by Chen et al.\cite{chen2008novel} to encapsulate lauric acid (LA) in PET. The resulted fibers were 710 nm in diameter and had around 71 J/g of latent heat. Nguyen et al.\cite{nguyen2011fabrication} fabricated PEG encapsulated in polyvinylidene fluoride (PVDF) fibers with an average diameter of around 700 nm in a form of non-woven mats. The fibers contained 20 \% by mass of PEG with 30-35 J/g of latent heat. Natural soy wax was encapsulated in polyurethane (PU) by Hu et al. \cite{hu2012encapsulation} in the shape of fibers with an average diameter from 0.65 to 1.8 $\mu$m and with up to 70 J/g of latent heat. In other two studies the same device was employed to encapsulate capric-lauric acid and capric-palmitic acid in PET having about 55 J/g of heat of fusion \cite{golestaneh2016fabrication}, as well as PEG in polyamide 6 (PA6) with maximum latent heat of 122 J/g \cite{babapoor2017coaxial}. Emulsion electrospinning slightly simplifies the process of encapsulation as does not require a coaxial flow. Such technique was applied by Zdraveva et al. \cite{zdraveva2015electrospun} to encapsulate plant oils that serve as PCM into Poly(vinyl alcohol). The fibers containing the PCM were between 400 and 700 nm in diameter depending on the PCM content and polymer concentration. The maximum latent heat achieved was about 97 J/g for the composite with the highest content of PCM. 

Despite the high storage capacity achieved by this method, the low speed of production \cite{cardenas2020systematic} and requirement of high voltage ranging from 12 to 20 kV \cite{nguyen2011fabrication,hu2012encapsulation} are important drawbacks of electrospinning. In order to overcome those limitations, a centrifugal or melt-spinning method can be used for PCM encapsulation into fibers. In the study of Yan et al. \cite{yan2021flexible} hollow polypropylene fibers were produced and later filled with PEG. With this approach fibers of about 1000 $\mu$m in diameter and with a very high PCM content on the order of 83 $\%$ by mass were reached that provided up to 130 J/g of latent heat. 

It was shown by Wen et al. \cite{wen2015microfluidic} and Zhang et al. \cite{zhang2018microfluidic} that encapsulation of PCMs into fibers was also possible by using microfluidic principles. A triple coaxial flow that consisted on RT27 (paraffin based PCM), dimethyl sulfoxide (DMSO) solution containing 14 \%  by mass of Poly(vinyl butyral) and an aqueous solution of carboxymethylcellulose sodium (1\% w/v) as inner, middle and outer fluids, respectively was created in a microfluidic device. The formation of the fiber sheath was driven by DMSO solvent extraction into water.  The maximum achieved content of RT27 was 70\% with a latent heat of 128.2 J/g. A similar approach but without microfluidic device and in two steps was accomplished by Xiang et al. \cite{xiang2019construction}. First, PVDF was dissolved in a poly(propylene carbonate) and dioctyl terephthalate mixture that afterwards was extruded into ethanol to extract the solvent and form hollow fibers. On the second step, the obtained fibers were filled with paraffin up to 77 \% by mass, providing around 80 J/g of latent heat. 

Despite quite high PCM content that can be achieved with some techniques, most of them are complex or require multiple steps to obtain the final product. Moreover, cost efficiency and possibility of upscaling for the different synthesis methods used so far is still questionable. Therefore, the aim of this paper is to present a simple, versatile, in-situ, cost-effective, easy to scale-up technology for the leakage free encapsulation of PCMs into flexible fibers with the main focus on thermal storage and thermal control applications at low temperature (e.g. building applications or battery thermal management system). In order to reach this goal, here the aforementioned above methods i.e. emulsion electrospinning, microfluidics and solvent extraction were combined. The morphology of the obtained encapsulated PCM was analyzed using SEM (Scanning Electron Microscope). In terms of thermal properties characterisation, DSC (Differential Scanning Calorimetry), TGA (Thermogravimetric Analysis) and DTG (Differential Thermogravimetric Analysis) were applied. Dynamic thermal performance of the fibers containing PCM during melting and crystallisation were visualised with the help of an infrared (IR) camera. Finally, tensile tests were carried out to evaluate mechanical properties of the fibers with and without PCM.

\begin{table*}  

\begin{framed}

\nomenclature[A]{$T$}{Temperature, [$^\circ$C]}
\nomenclature[A]{$E_{en}$}{Encapsulation efficiency}

\nomenclature[B]{$SD$}{Standard deviation}
\nomenclature[B]{$EEn$}{Encapsulation efficiency}
\nomenclature[B]{$PCM$}{Phase change material}
\nomenclature[B]{$TES$}{Thermal energy storage}
\nomenclature[B]{$PDC$}{Pad-dry-cure (PDC) method}
\nomenclature[B]{$PEG$}{Polyethylene glycol}
\nomenclature[B]{$PET$}{Polyethylene terephthalate}
\nomenclature[B]{$LA$}{Lauric acid}
\nomenclature[B]{$PU$}{Polyurethane}
\nomenclature[B]{$DMSO$}{Dimethyl sulfoxide}
\nomenclature[B]{$CMC$}{carboxymethylcellulose sodium}
\nomenclature[B]{$PVDF$}{Polyvinylidene fluoride}
\nomenclature[B]{$SEM$}{Scanning Electron Microscope}
\nomenclature[B]{$DSC$}{Differential Scanning Calorimetry}
\nomenclature[B]{$TGA$}{Thermogravimetric Analysis}
\nomenclature[B]{$DTG$}{Differential Thermogravimetric Analysis}
\nomenclature[B]{$IR$}{Infrared}
\nomenclature[B]{$DMF$}{N,N-Dimethylformamide}
\nomenclature[B]{$BSED$}{Backscattered electron detector}

\nomenclature[O]{$\Delta H$}{Latent heat, [J/g]}

\nomenclature[N]{$1,2,3$}{Correspond to peak temperature sequence}

\printnomenclature
\end{framed}

\end{table*}

\section{Experimental section}

\subsection{Materials}
Paraffin RT-28HC with a melting point between 27 and 29 ºC and a latent heat of 250 J/g was supplied by Rubitherm® \cite{Rubitherm} and used as core material of the composite fibres. Commercial PVDF Solef® 5130 from SOLVAY was dissolved in N,N-Dimethylformamide (DMF) (99.5\% Pure) from EMPARTA®. This solution was used as a precursor for fiber production. All chemicals were used as received with now further purification. Tap water was utilized for the DMF extraction process.

\subsection{Preparation of the precursors}
Two kind of precursor solutions were prepared for the production of pure PVDF and composite fibres. The precursor for the pure PVDF fibre consisted of a 5\% by mass solution of PVDF in DMF that was prepared by dissolving PVDF in DMF at 60 ºC with continuous stirring. As a precursor of the composite fibre, the same 5\% PVDF solution was used in which 5\% by mass of paraffin was added and emulsified by sonication in an ultrasound bath (S60H Elmasonic) at 40 $^{\circ}$C for 30 minutes. No surfactant was used to stabilize the emulsion.

\subsection{Production of the fibres}
\begin{figure}
\centering
\includegraphics[width=0.70\linewidth]{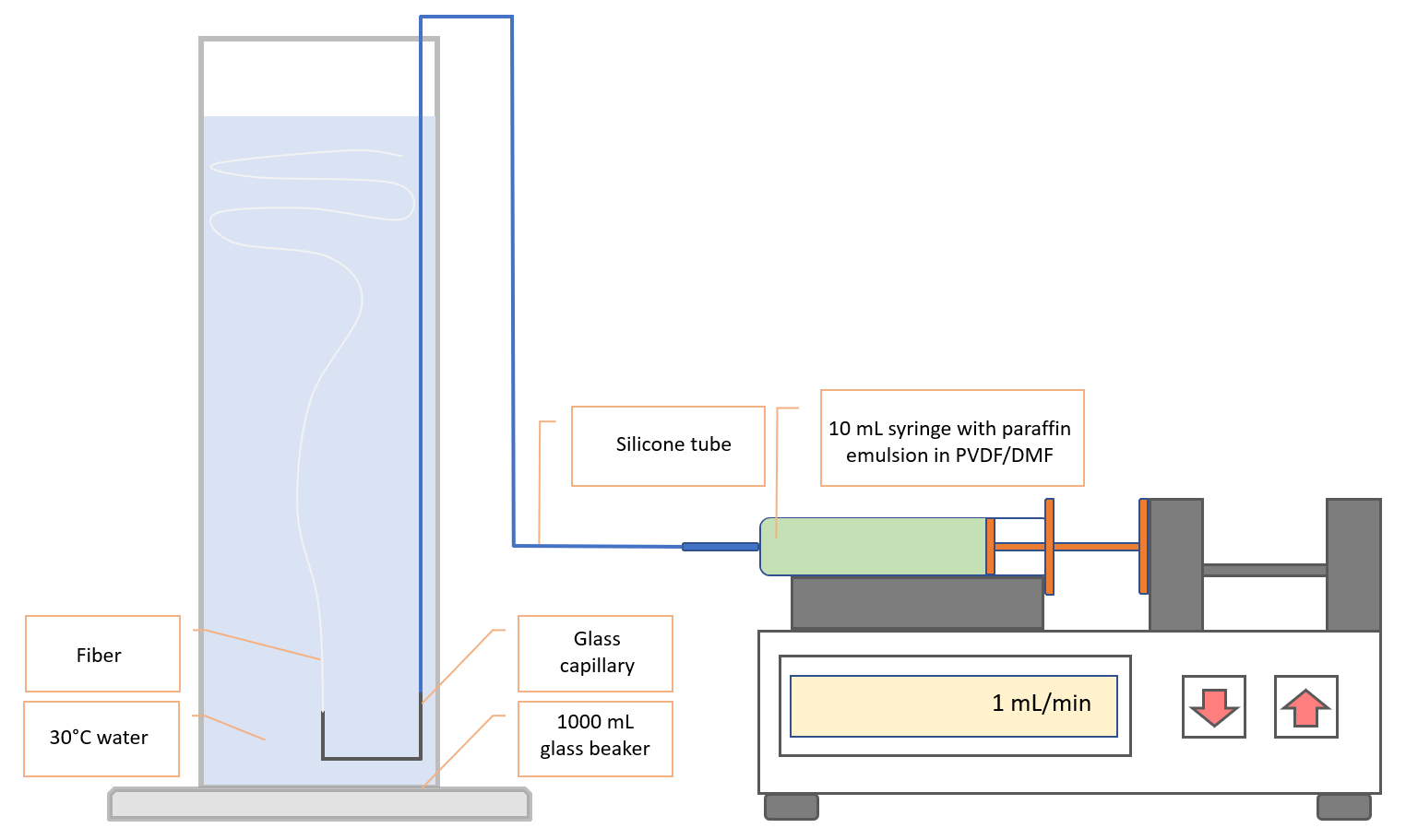}%
\caption{Scheme of the experimental setup  for fibers production}
\label{fig:Exp setup}
\end{figure}

In this study, the microfluidic approach of coaxial flows \cite{wen2015microfluidic,zhang2018microfluidic} was further simplified to gravity aided laminar jet injection into a bulk fluid, to produce fibers.
A scheme of the experimental setup is shown in Figure \ref{fig:Exp setup}. The fibre precursors were injected from a glass capillary (inner diameter of 0.47 mm) into a tall glass beaker containing tap water at 30 $^{\circ}$C. Immediately after injection, the solvent extraction process took place, extracting DMF to the tap water. In this way, the formation of the fiber sheath takes place trapping the paraffin inside.

\begin{figure}
\centering
\includegraphics[width=1.00\linewidth]{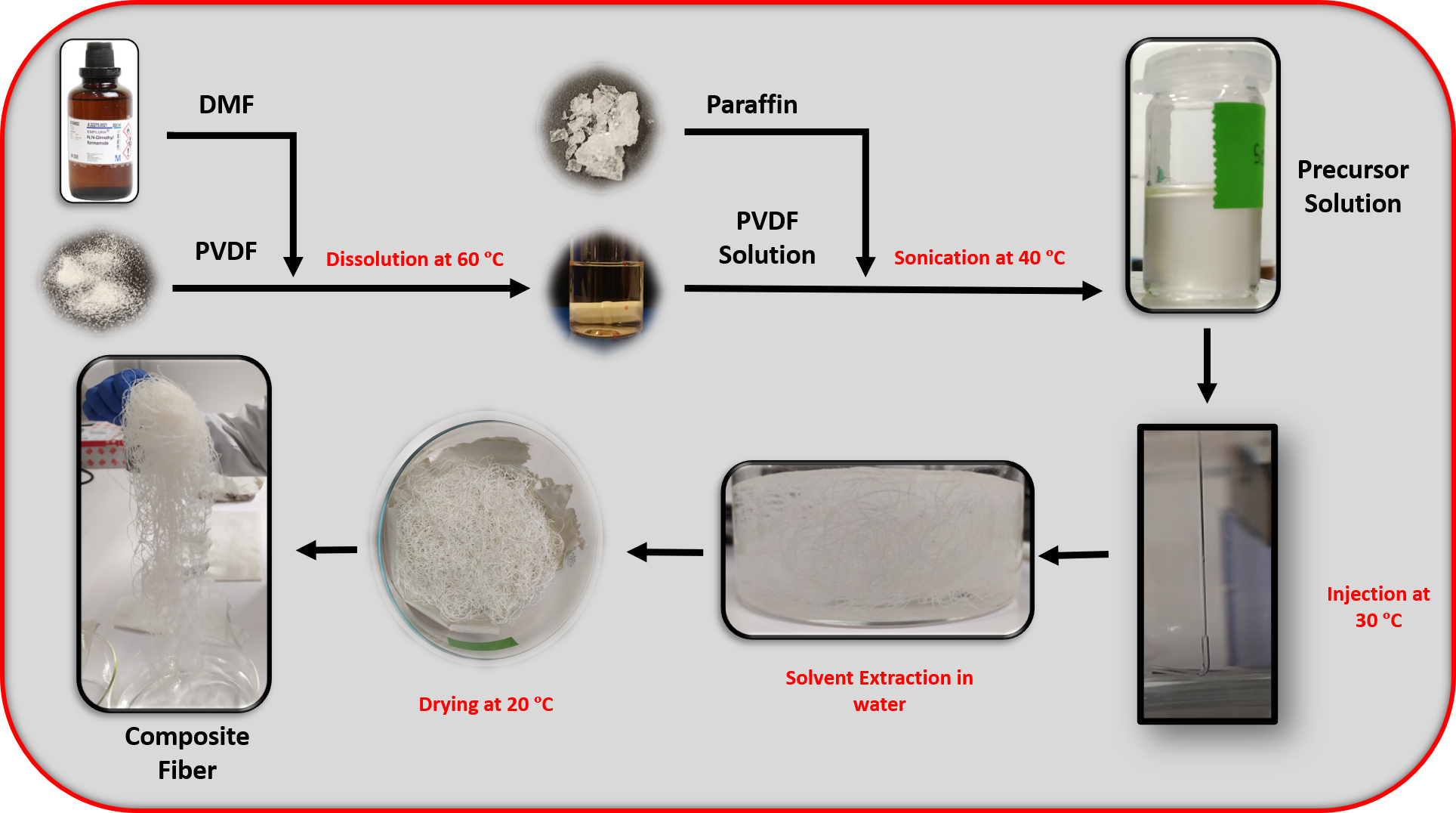}%
\caption{Schematic representation of the production procedure}
\label{fig:Production scheme}
\end{figure}

The injection process was started less than 10 minutes after emulsion preparation to avoid coagulation of the paraffin droplets. A syringe  pump (Pump 11 - Pico Plus Elite from HA Harvard Apparatus) was used to support the flow of precursor (at flow rate of 0.5 mL/min) from the syringe to the glass capillary that were connected by a silicone tube. The use of a syringe pump is not obligatory, as  preliminary tests with a manually driven syringe showed rather good results. However, the use of the syringe pump helps to achieve fibers with constant diameter, witch can be important depending on the final application. To avoid paraffin solidification during all the fibre production process, the glass beaker and the air surrounding the experimental setup were heated up to 30 $^{\circ}$C by an electric fan-heater. After the injection process, the fibres were collected and stored in tap water for 24h for the maximum extraction of DMF and then were dried during 48h at 20 $^{\circ}$C in a fume hood. Schematic representation of the production procedure is shown in Figure \ref{fig:Production scheme}.

\subsection{Characterization}
\subsubsection{Scanning electron microscopy}
To analyze the microstructure of the fibres, samples were imaged
by means of a scanning electron microscope Quanta 200 FEG operated in
low vacuum mode at 10 kV featured with a backscattered electron detector
(BSED). The operational
conditions were carefully selected to avoid the melting of the
paraffin during the study.

\subsubsection{DSC analysis}
DSC was used to determine the  temperatures of phase transition and their corresponding latent heats. DSC was used as well to measure the specific heat of fibres without paraffin, original PVDF powder and paraffin used for precursors preparation. A power-compensation DSC Q2500 from TA Instruments was employed with sealed aluminum crucibles. The mass of the samples was ca. 9 mg. Argon (50 ml/min) was employed as purge gas. Each sample was subjected to three cycles of heating and cooling, with a heating/cooling rate of 5 $^\circ$C/min. The transition temperatures were determined by the onset temperature of the corresponding endothermic peaks in the DSC-thermograms, whereas the latent heat was calculated by integration of the latter assuming a linear baseline. The DSC was calibrated for heat flow and temperature using high purity (>99.99\%) reference materials indium and sapphire. The accuracy in the determination of melting temperatures is ±0.5 $^\circ$C, whereas that for the enthalpy of phase transition and specific heat is ±5\%.

\subsubsection{Thermal stability}
Thermogravimetric analysis of the fibres and raw materials was performed in a TG209F1 Libra® Thermogravimetric Analyzer from NETZSCH. The analysis was based on a heating ramp from room temperature to 600 $^\circ$C with a heating rate of 10 $^\circ$C/min under nitrogen atmosphere.

\subsubsection{Leakage test} \label{Leakage test}
A leakage test was carried out to check the PCM retention capacity of the composite fiber. Thus, a piece of composite fiber of 200 mg was subjected to 1000 melting/solidification cycles. The fiber was placed onto a Peltier element, covered with absorbing paper and then pressed with an aluminum block of 61 g (40x40x13 mm) to ensure a homogeneous temperature distribution and good contact between the fiber and the absorbing paper. Thermal cycles were performed from 10 to 40 $^\circ$C. Isotherms of 120 seconds were held at maximum and minimum temperatures for each cycle. Thermal regulation of the Peltier element was ensured using a type K thermocouple and an Arduino Nano coupled with a relay board (Parallax Inc 27115) and a MAX 31856 module (thermocouple amplifier/converter with integrated cold-junction compensation). Accuracy of the measurement of the thermocouple module was checked at different temperatures using a temperature-controlled bath (Julabo 1000F) coupled with other systems of temperature measurement (National Instruments NI 9211 and Fluke 233). According to these tests, the accuracy of temperature measurements is less than 2\% of a given reading. A software dedicated to the thermal regulation and data acquisition was coded with Python 3.7.3, Arduino 1.8.13 and using the Adafruit\_MAX31856 library \cite{GitHub}. The mass of the fiber was checked using a Mettler Toledo ML304T/00 balance with the accuracy of 0.4 mg at the beginning of the test and after 50, 100 and 1000 cycles to quantify PCM leakage.

\subsubsection{Thermal performance} \label{Thermal performance}
To study the thermal behavior of the manufactured fibers at a larger scale, 4 samples of pure PVDF and 4 of composite fibers were placed in parallel over a Peltier element and subjected to heating/cooling cycles from 10 to 40 $^\circ$C (see description of the thermal regulation in section \ref{Leakage test}). The thermal behavior of these samples was recorded during the cycles using an infrared camera (FLIR A6752sc). The camera was configured for data acquisition at 30 Hz in a calibrated temperature range of -20 $^\circ$C to 55 $^\circ$C (integration time was fixed at 3.3 ms). Conversion of the radiometric data to apparent temperatures was performed using the internal calibration curves of the camera and fixing the emissivity of the set of fibers at 0.8 \cite{Emisivity}. Reflected and atmospheric temperatures were considered constant (20 $^\circ$C). 
After cycling, processing of the recorded data was performed using the FLIR software ResearchIR Max Version 4.40.11.35. Firstly, apparent temperature of each fiber was calculated by averaging the apparent temperatures of a pixel’s line drawn along the length of each of them. Then, to visualize more easily the localization of the phase change, a simple image processing based on the subtraction of the tenth image from the current frame (sliding subtraction) was performed.

\subsubsection{Mechanical properties}
Tensile stress and strain of the fibres were measured using an Instron 34SC-5 single column universal testing machine with a 100 N capacity sensor. Single fibers were placed between the machine clamps with the help of paper "supports/holders". The measurements were carried out at room temperature with a fixed initial gauge length of 20 mm and a stretching rate of 10 mm/min. Measurement accuracy of the column is ±0.5\% for load and ±0.02 mm or 0.15\% of total displacement (whichever is greater) for strain values. To obtain an average value, ten samples of each fibre (pure PVDF and composite fibres) were tested.

\section{Results and discussions}

\subsection{Morphology} \label{Sec.Morphology}
\begin{figure}
\centering
\begin{minipage}[h]{0.30\linewidth}
\center{\includegraphics[width=1\linewidth]{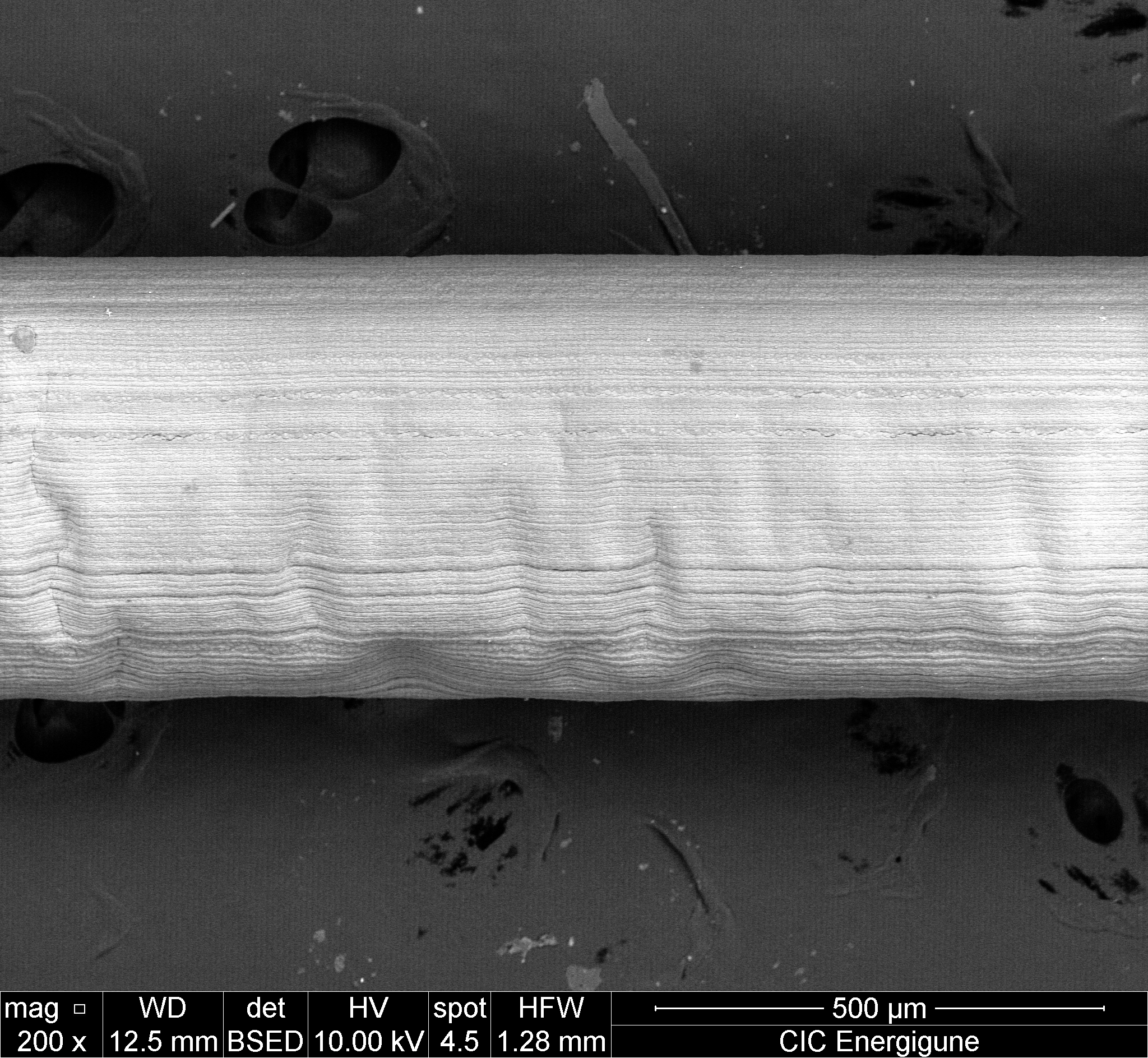} \\ (a)}
\end{minipage}
\begin{minipage}[h]{0.30\linewidth}
\center{\includegraphics[width=1\linewidth]{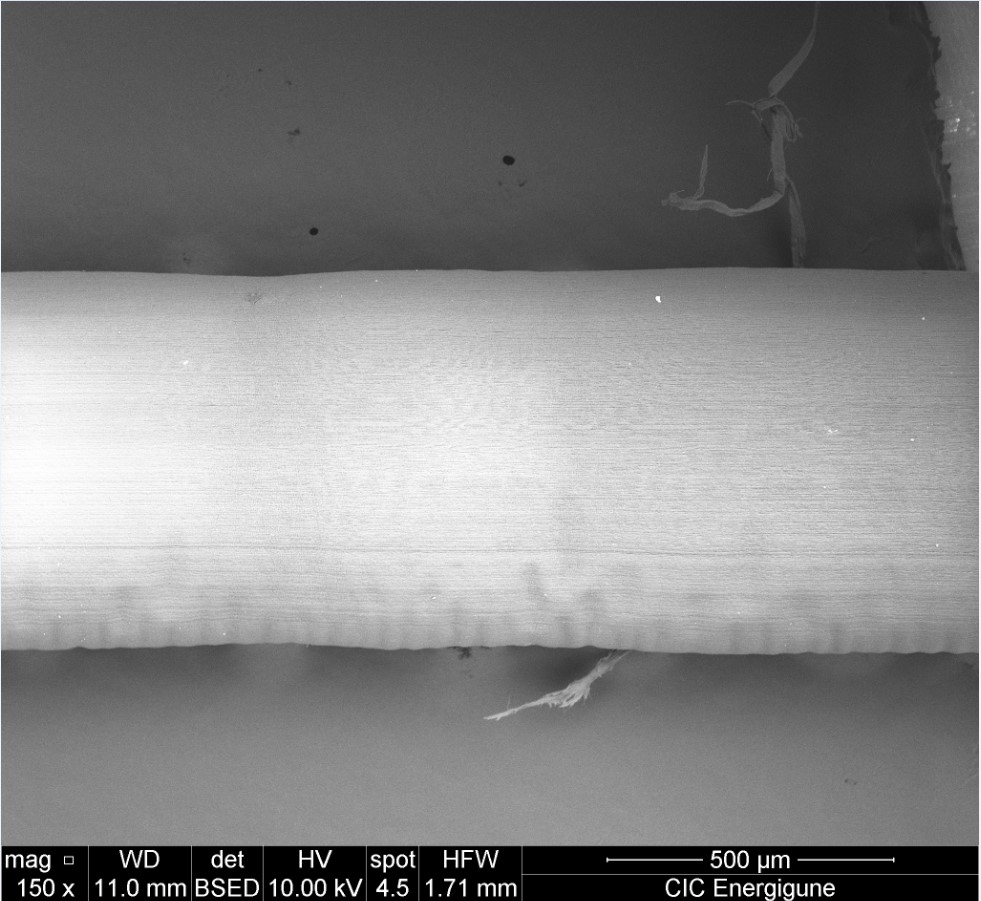} \\ (e)}
\end{minipage}

\begin{minipage}[h]{0.30\linewidth}
\center{\includegraphics[width=1\linewidth]{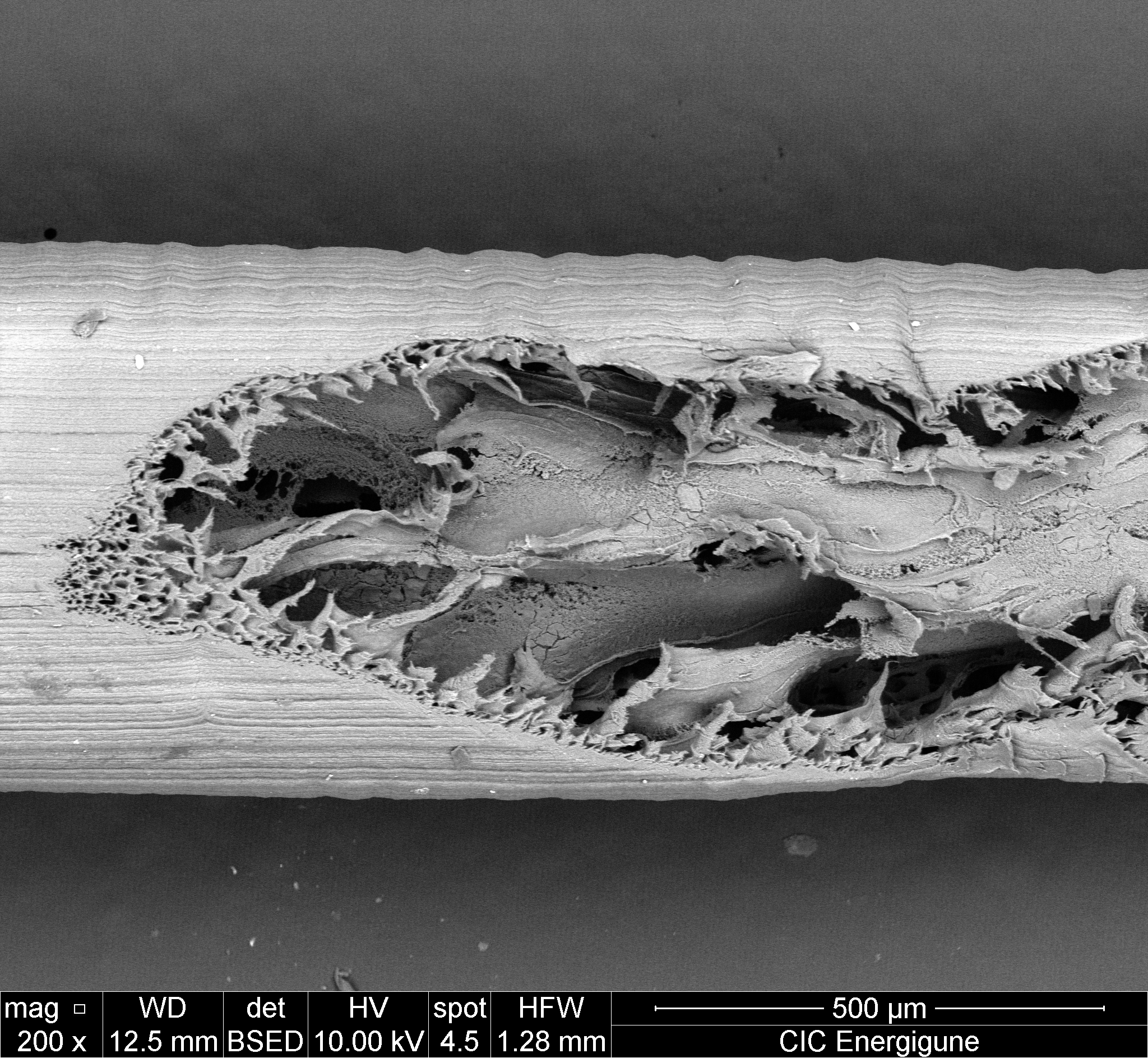} \\ (b)}
\end{minipage}
\begin{minipage}[h]{0.30\linewidth}
\center{\includegraphics[width=1\linewidth]{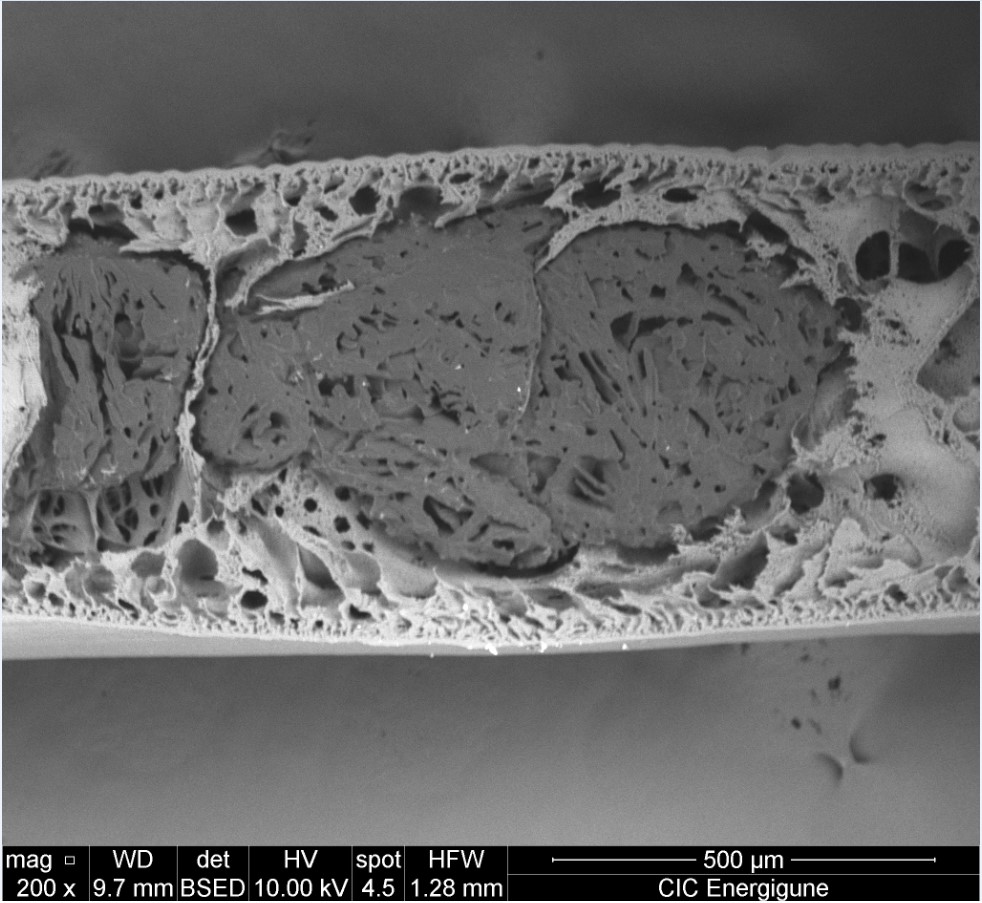} \\ (f)}
\end{minipage}

\begin{minipage}[h]{0.30\linewidth}
\center{\includegraphics[width=1\linewidth]{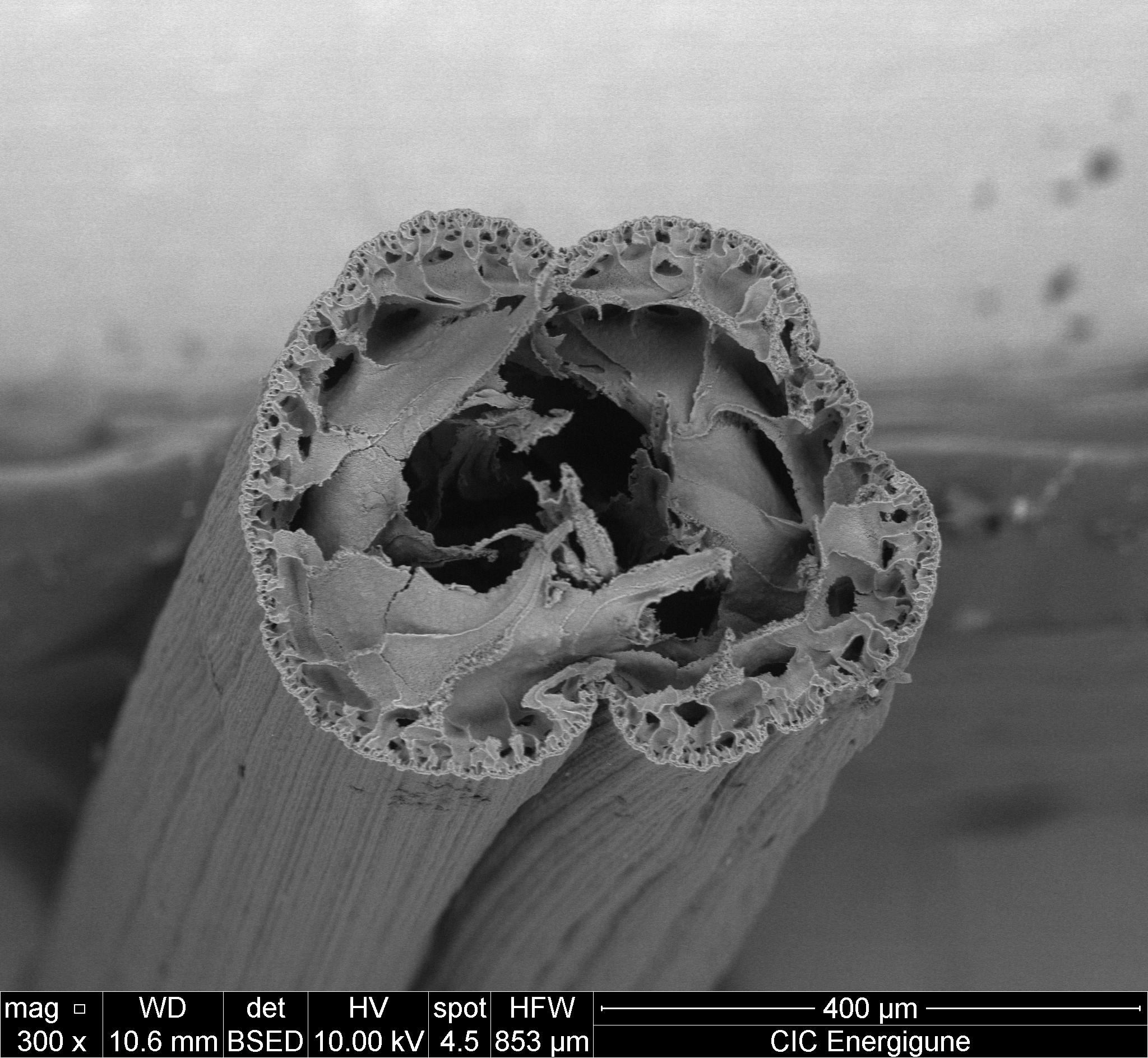} \\ (c)}
\end{minipage}
\begin{minipage}[h]{0.30\linewidth}
\center{\includegraphics[width=1\linewidth]{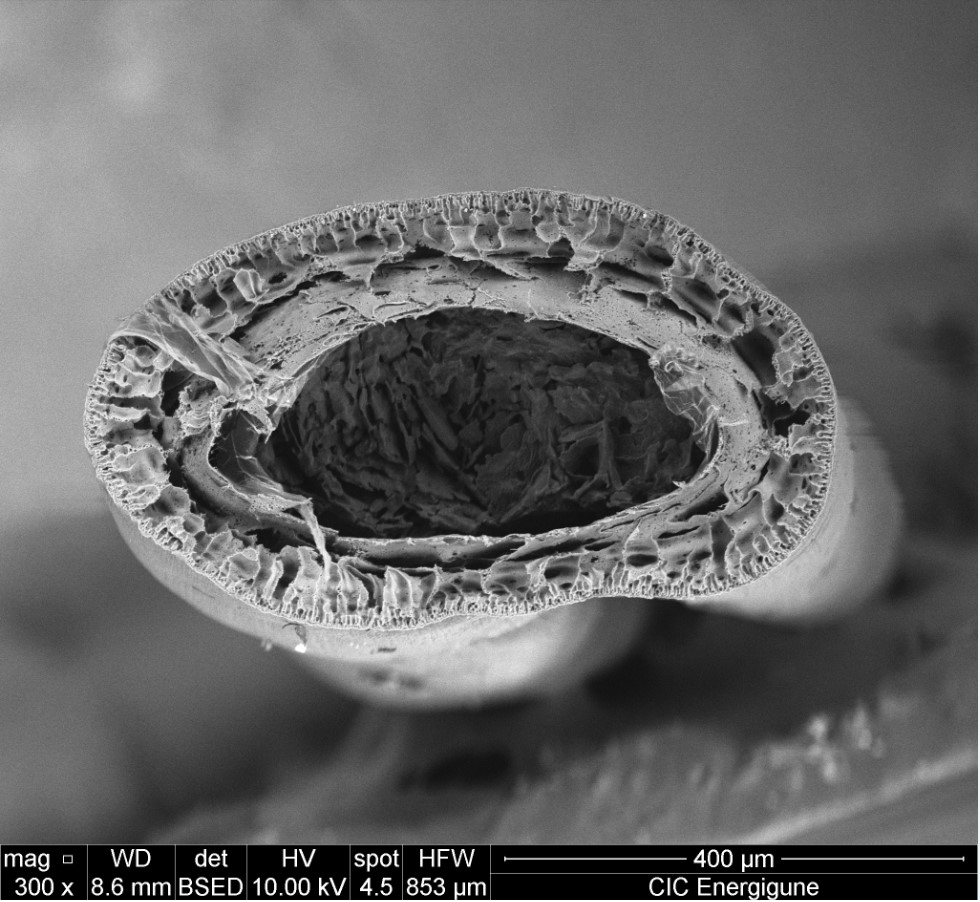} \\ (g)}
\end{minipage}

\begin{minipage}[h]{0.30\linewidth}
\center{\includegraphics[width=1\linewidth]{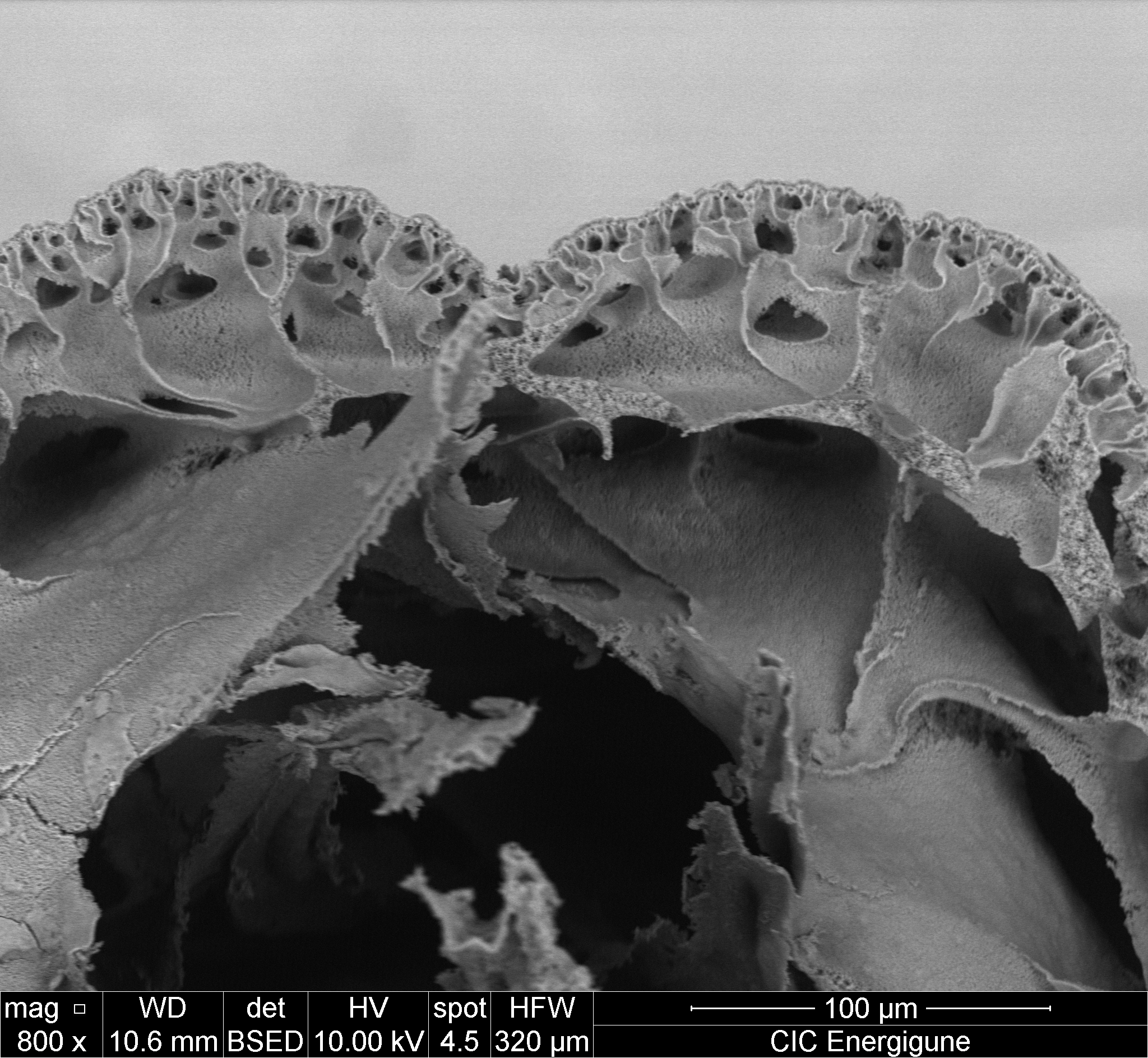} \\ (d)}
\end{minipage}
\begin{minipage}[h]{0.30\linewidth}
\center{\includegraphics[width=1\linewidth]{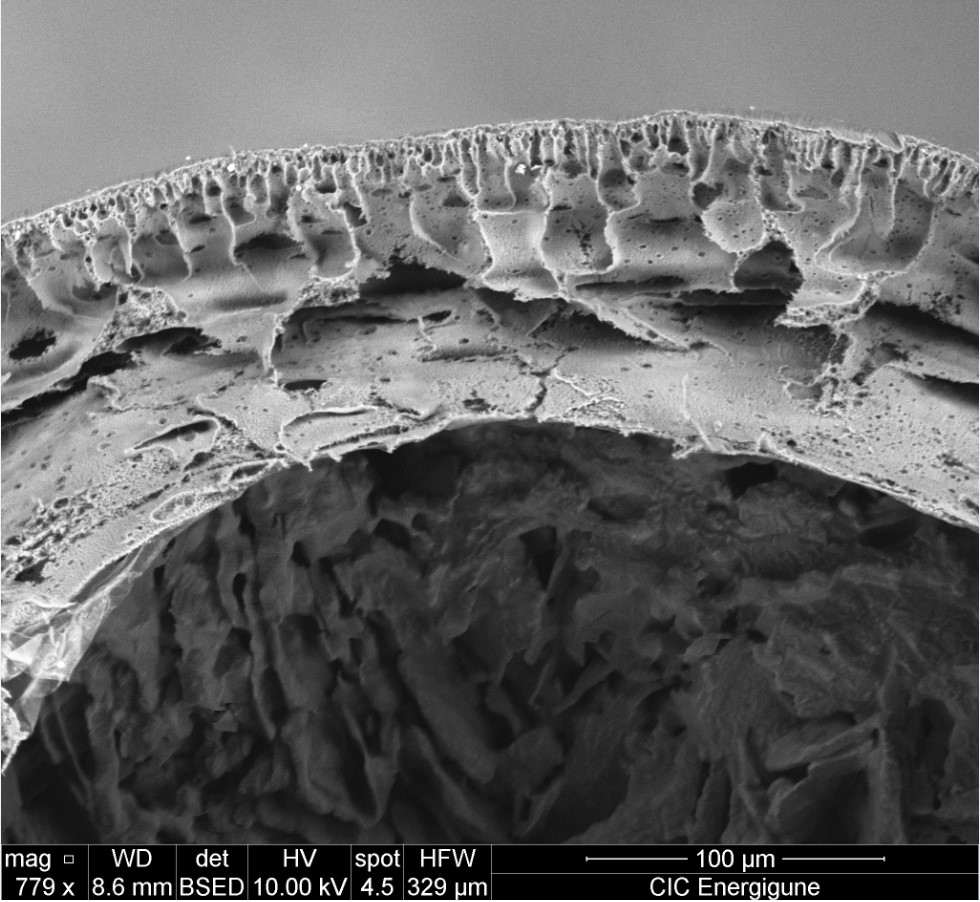} \\ (h)}
\end{minipage}

\caption{SEM images of PVDF and PVDF/PCM fibers: (a,e) side views; (b,f) side cuts; (c,d and g,h) cross cuts }\label{SEM}
\end{figure}

The cross cut of the fibers was done by stretching until break-up of a piece of the sample immersed in the liquid nitrogen with the help of two tweezers. The side cut was made by scalpel at $\sim$20 $^{\circ}$C. SEM images of surface, cross section and axial section both for pure PVDF and composite fibers are presented in Figure \ref{SEM}. First, one can see that the outer diameter of the produced fibers is very close to that of the used for injection capillary that is around 0.5 mm. The wall thickness of the fibers is about 0.1-0.2 mm. The exterior surface of both fibers, with and without paraffin consist of a dense PVDF without imperfections or holes. Moreover, pure PVDF fibers are covered by longitudinal folds, while the surface of the composite fibers looks flatter (see Figure \ref{SEM} c,d and e,h). Those folds are a consequence of the solvent extrication and drying processes. In the case of composite fibers, the paraffin play the role of support, as well as suppress solvent extraction rate that finally result in smoother surface.

The average cross sectional effective density of the PVDF fibers produced by solvent extraction has its maximum at the fiber's wall and decrease until zero or near zero value at the central axe. Moreover, fibers are porous with three‐modal pores size distribution. The largest macro pore "channel" on the order of 50-300 $\mu$m is located in the middle of the fiber. Fiber's walls are also porous and have two characteristic size ranges of 1-50 $\mu$m and the other less than 1 $\mu$m. 


The composite fibers have a core-sheath structure where the paraffin (core) is wrapped by PVDF porous shell. The outer wall is smooth, dense and homogeneous, presenting no pores, which indicates that it forms a sealed container that will help to avoid PCM leakage. From cross-sectional images (Figure \ref{SEM} (g,h)) a continuous paraffin core along the fiber could be assumed but the axial-cut images (Figure \ref{SEM} (f)) show that the paraffin is non homogeneously distributed in the form of “slugs”. These “slugs” differ in size and are separated by PVDF walls along all the fiber length, that helps to avoid the leakage of all the paraffin in the case of punctual ruptures of the fibre. 

The morphology of produced fibers is similar to that reported in several studies \cite{xiang2019construction,wen2015microfluidic,zhang2018microfluidic} and as was shown by Grosu at al. \cite{grosu2020hierarchical} such hierarchical porous structures are providing outstanding anti-leakage properties.

\subsection{Thermal Properties}
\subsubsection{DSC Analysis} \label{Sec.DSC}

\begin{figure}
\centering
\includegraphics[width=0.50\linewidth]{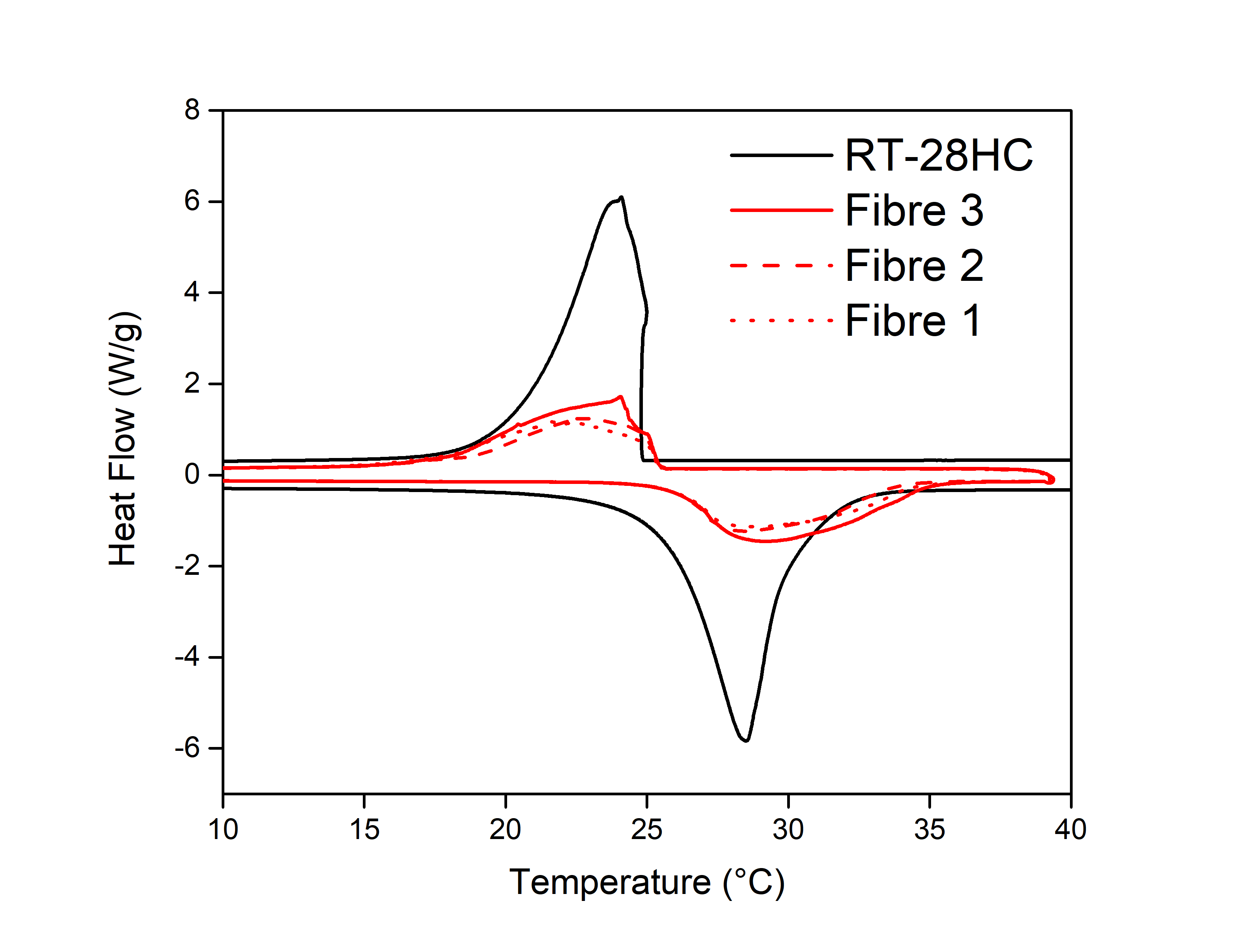}%
\caption{DSC thermograms of pure paraffin and different samples from the obtained composite fibre.}
\label{fig:DSC}
\end{figure}

The temperatures of phase transition and corresponding latent heat of paraffin (RT-28HC) and the synthesised composite PVDF/paraffin fibres were measured by DSC. Each sample was submitted to three heating and cooling cycles. Figure \ref{fig:DSC} shows the obtained thermograms during the third cycle for both pure PCM and three randomly selected composite fibre samples. For the composite fibres, thermograms of three randomly selected samples from the same bulk are shown. Onset temperatures, latent heat values and calculated encapsulation efficiencies are given in Table \ref{Tab:DSC}. From the DSC analysis the encapsulation efficiencies can be obtained according to the following equation:

\begin{equation}
    E_{en}=\frac{\bigtriangleup H_{m\_Fibre}}{\bigtriangleup H_{m\_Paraffin}}\cdot100
    \label{eqn:encapsulation}
\end{equation}

where, $H_{m\_Fibre}$ and $H_{m\_Paraffin}$ are the latent heat of composite fibers and paraffin respectively.

As shown in Figure \ref{fig:DSC}, both paraffin and the composite fibres present a single endothermic peak during heating and cooling processes. The melting starts at 25.7 $^{\circ}$C for paraffin and at 26.2 $^{\circ}$C for all the composite fibre samples. On the other hand, the peak appears at 24.7 $^{\circ}$C for the pure sample and at 25.3 $^{\circ}$C for the composite fibres during cooling cycle because of the recrystallization of paraffin.

It can be seen that crystallization onset temperature of the paraffin is higher when encapsulated into fibers. It is a confinement effect of the fiber that promotes paraffin crystallization and reduces slightly its subcooling.

All tested samples have shown latent heat values of melting between 73 and 98 J/g. Comparing to the latent heat of pure paraffin (247 J/g) and using the Equation \ref{eqn:encapsulation}, the corresponding encapsulation efficiency lay in the range from 29 to 39\%. This variation of the encapsulation efficiency of randomly chosen samples is obviously related to the inhomogenous distribution of the paraffin alongside the composite fibres. Even though, the emulsion was stable during composite fibers production, at the stage of solvent extraction a strong gradients of PVDF and DMF force coalescence of paraffin droplets that cause formation of larger paraffin fractions. Finally it results in such inhomogeneity that can be seen on the SEM images (Figure \ref{SEM}) and will be shown later in the Section \ref{Sec.IR}, where thermal response of the composite fibers was captured by IR camera.

\begin{table}[]
\caption{Transition temperatures, related latent heat and encapsulation efficiency values of the paraffin and composite fibres}
\begin{tabular}{lccccccc}
\textbf{Sample}  & \textbf{Tm,o ($^\circ$C)} & \textbf{Tm,p ($^\circ$C)} & \textbf{$\Delta$Hm (J/g)} & \textbf{Tc,o ($^\circ$C)} & \textbf{Tc,p ($^\circ$C)} & \textbf{$\Delta$Hc (J/g)} & \textbf{Een (\%)} \\
\textbf{RT-28HC} & 25.7              & 28.5              & 247             & 24.7              & 24.1              & 247             & -                 \\
\textbf{Fibre 1} & 26.2              & 28.5              & 75             & 25.3              & 22.4               & 75             & 30          \\
\textbf{Fibre 2} & 26.2              & 28.5              & 73             & 25.3              & 23.2              & 72             & 29          \\
\textbf{Fibre 3} & 26.2              & 29.2              & 98             & 25.4              & 24.1              & 97             & 39         
\end{tabular}
\label{Tab:DSC}
\end{table}

\begin{figure}
\centering
\includegraphics[width=0.50\linewidth]{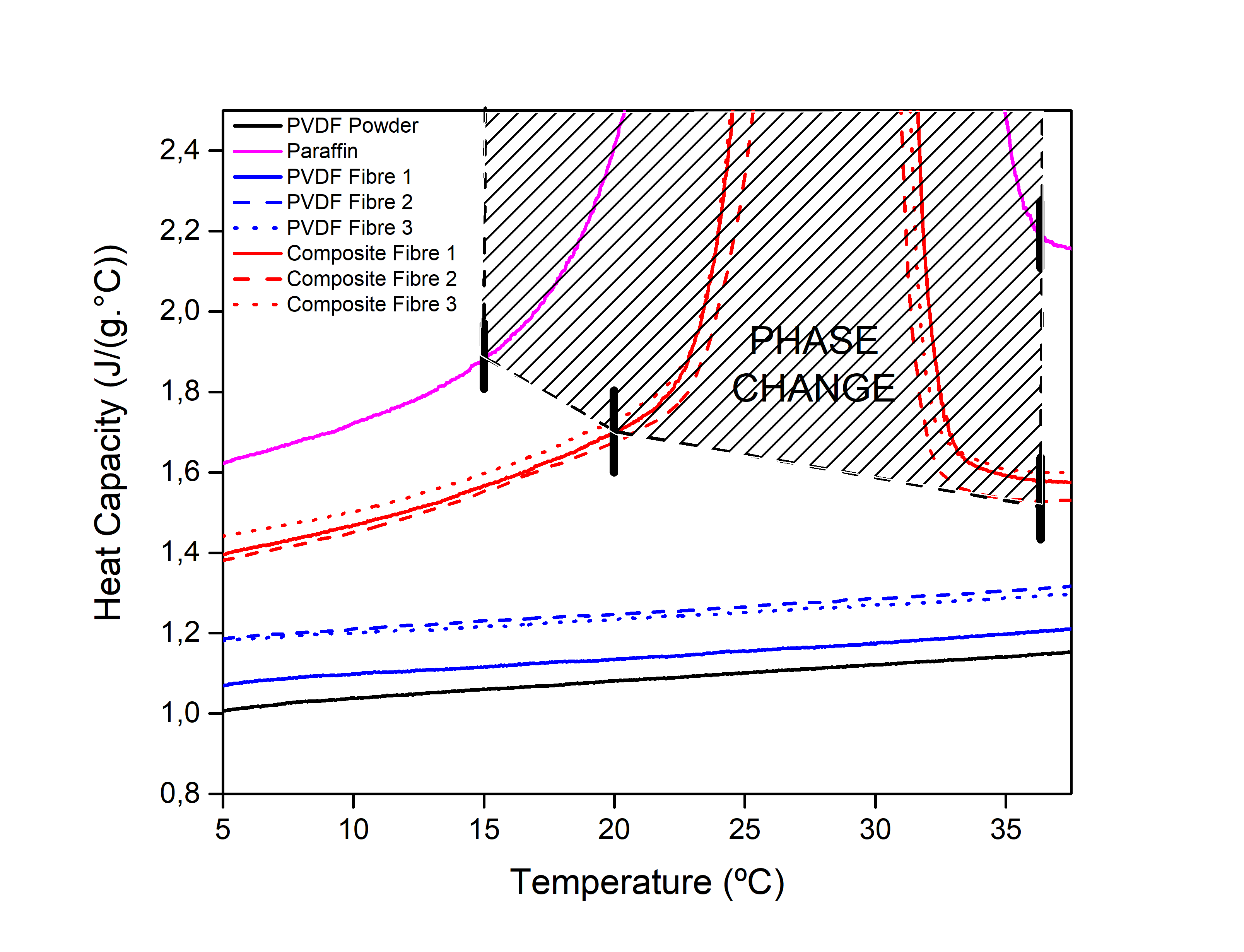}%
\caption{Cp of the obtained fibres and their components.}
\label{fig:Cp}
\end{figure}

Moreover, Figure \ref{fig:Cp} and Table \ref{Tab:Cp} indicate the specific heat profiles and values obtained from DSC analysis of the fibres and the components it composed of. It can be seen that all the PVDF fibres present higher Cp values than that for pure PVDF powder while temperature dependence of Cp is the same. The higher Cp values of PVDF fibres reveal their lower degree of crystallinity compared to the supplied PVDF powder \cite{bioki2012effect}. Moreover, the Cp of the composite fibre is even higher, reaching values between 1.53 and 1.60 J/g$\cdot$K at 37.5 $^\circ$C due to the presence of paraffin, that has higher Cp values.

\begin{table}[]
\caption{Cp Values for the produced fibres an its pure components}
\begin{tabular}{ccccc}
\multirow{2}{*}{Sample} & \multicolumn{2}{c}{Solid state}             & \multicolumn{2}{c}{Liquid state} \\
                        & Temperature range ($^\circ$C) & Equation (J/g·$^\circ$C)      & Temperature ($^\circ$C)   & Value (J/g·$^\circ$C)    \\
PVDF Powder             & 5 - 37.5                         & 0.995+0.004*T & ---                    & ---     \\
PVDF Fibre 1            & 5 - 37.5                          & 1.056+0.004*T & ---                    & ---     \\
PVDF Fibre 2            & 5 - 37.5                         & 1.171+0.004*T & ---                    & ---     \\
PVDF Fibre 3            & 5 - 37.5                          & 1.165+0.004*T & ---                    & ---     \\
RT-28HC                 & 5 - 15                         & 1.486+0.024*T & 37.5                   & 2.156   \\
Composite Fiber 1       & 5 - 20                         & 1.277+0.020*T & 37.5                   & 1.576   \\
Composite Fiber 2       & 5 - 20                         & 1.262+0.020*T & 37.5                   & 1.530   \\
Composite Fiber 3       & 5 - 20                        & 1.322+0.019*T & 37.5                   & 1.603  
\end{tabular}
\label{Tab:Cp}
\end{table}

\subsubsection{Thermal Stability (TGA)} \label{Sec.TGA} 

\begin{figure}
\centering
\begin{minipage}[h]{0.49\linewidth}
\center{\includegraphics[width=1\linewidth]{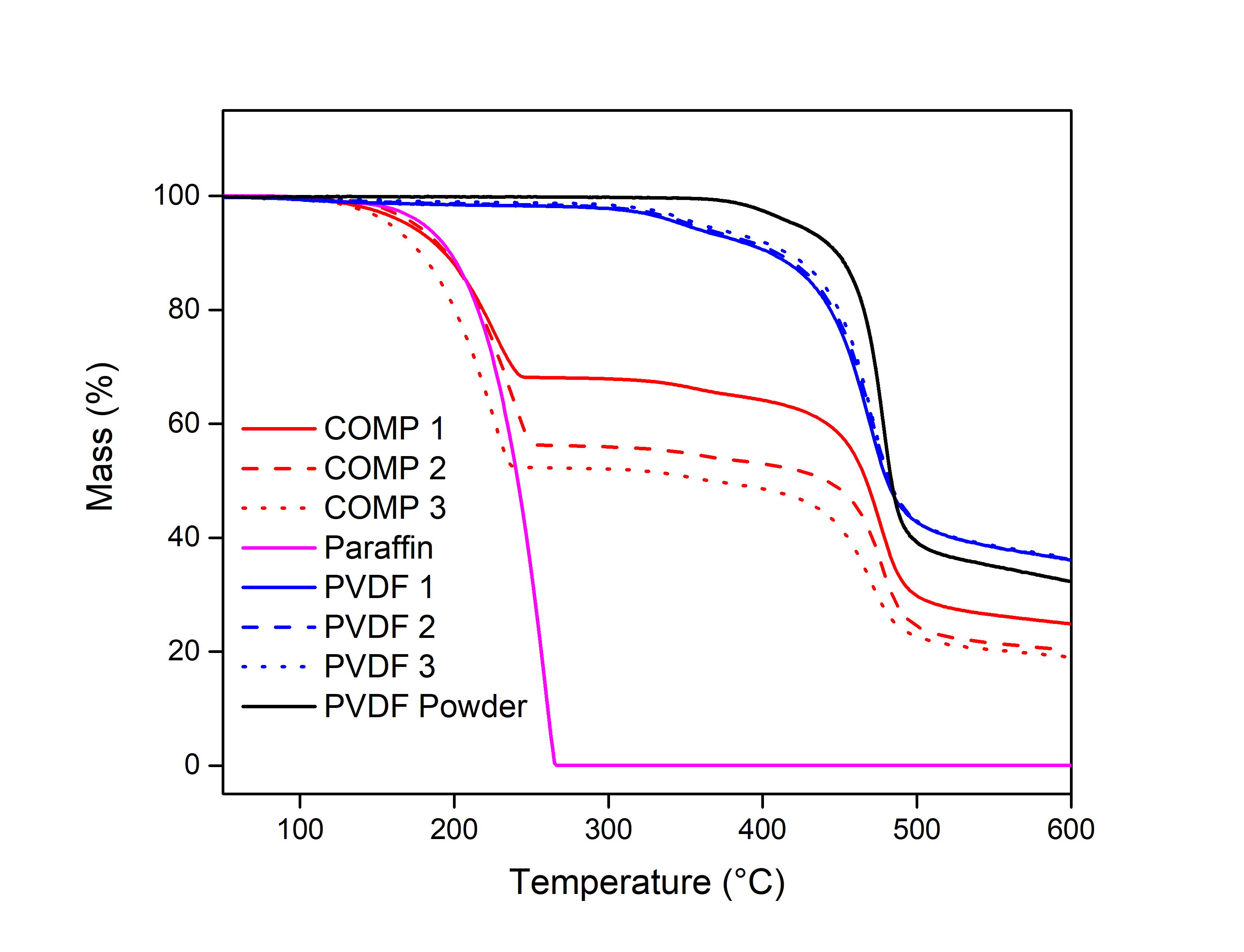} \\ (a)}
\end{minipage}
\begin{minipage}[h]{0.49\linewidth}
\center{\includegraphics[width=1\linewidth]{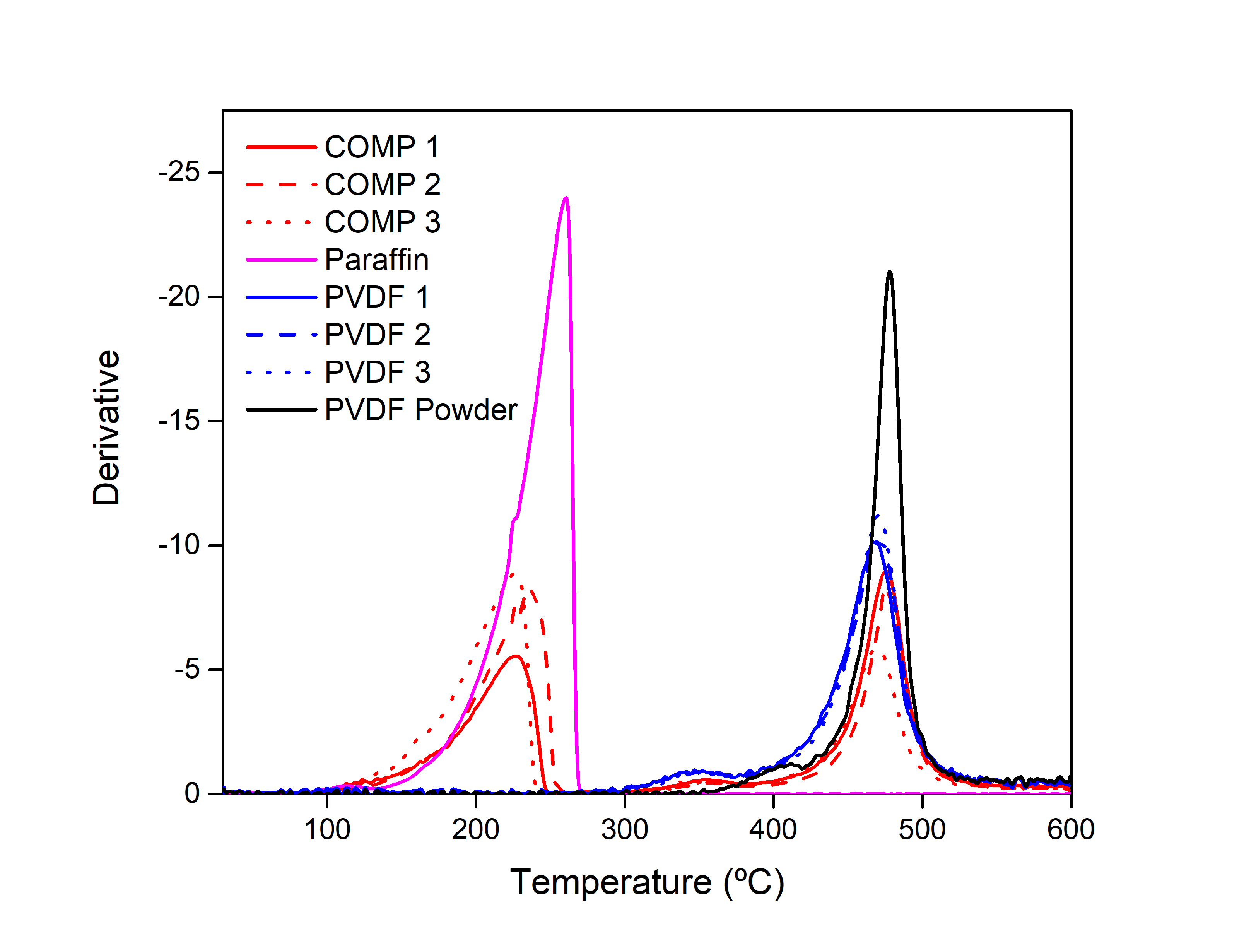} \\ (b)}
\end{minipage}
\caption{TGA (a) and DTG (b) profiles of pure paraffin and the manufactured PVDF-hollow and composite fibres}\label{TGA/DGT}
\end{figure}

The TGA and DTG profiles of pure paraffin, PVDF powder, PVDF fibre and the composite fibre are presented in Figure \ref{TGA/DGT}, while the degradation temperature, mass loss percentage and final residue quantities of each sample are shown in Table \ref{Tab:TGA}.

As can be seen from Figure \ref{TGA/DGT}, paraffin RT-28HC is thermally stable until 179.8 $^\circ$C. Then thermal degradation (it is evaporation in fact) starts and reaches its maximum rate at 224.5 $^\circ$C with no residue at the end of the process. Pure PVDF powder presents good thermal stability until 360 $^\circ$C. Then follow two degradation peaks, lower one at 411 $^\circ$C and the maximum one at 478.2 $^\circ$C.

Similar to the pure PVDF powder, two main degradation peaks can be observed for the PVDF fibres. The onset of the degradation takes place at 300 $^\circ$C, that is 53.3 $^\circ$C lower than that for PVDF powder. The first minor peak appears at 349 $^\circ$C and second major one at 469.6 $^\circ$C. As one can see, the onset and both degradation peaks are shifted to the lower temperatures for PVDF fibres compared to PVDF powder. This indicates the change in crystallinity of the PVDF due to the rapid crystallisation by solvent extraction method \cite{shi2012effect}.


Finally, the composite fibres present three main degradation peaks. The first one appears at an average temperature of 230.8 $^\circ$C and is related to the paraffin content inside the fibres, that varies from 31.8 to 47.5\%. This variation is a consequence of the already mentioned non homogeneous distribution of the paraffin inside the fibres, that could be seen in SEM images and that also affected the encapsulation efficiency variations of the DSC analysis results.

It has to be mentioned that this average decomposition temperature of the encapsulated paraffin is 6 $^\circ$C higher than that obtained for the pure PCM, which indicates that the PVDF sheath of the fibres is protecting the PCM from thermal degradation (evaporation). The onset of the PVDF degradation appears at 300 $^\circ$C similar to that of PVDF fibers. However, the second peak occurs at 353.3 $^\circ$C and the third one at 473.3 $^\circ$C. Both peaks corresponding to PVDF degradation are at $\sim$4 $^\circ$C higher than for pure PVDF fibers. This is in coherence with the results on specific heat of produced fibers and confirms higher crystallinity of PVDF in composite fibers.

\begin{table}[]
\caption{Degradation onset and peak temperature, mass loss percentage and residual mass of the samples}
\begin{tabular}{ccccccccc}
\textbf{Sample}      & \textbf{T(-5\%)}     & \textbf{T1 ($^\circ$C)}     & \textbf{Loss 1 (\%)} & \textbf{T2 ($^\circ$C)}     & \textbf{Loss 2 (\%)} & \textbf{T3 ($^\circ$C)}     & \textbf{Loss 3 (\%)} & \textbf{Residue (\%)} \\
\textbf{Comp.1}      & 170.1                & 227.7                & 31.8                 & 353.2                & 3.1                  & 475.9                & 39.3                 & 24.9                  \\
\textbf{Comp.2}      & 175.0                  & 238.1                & 43.8                 & 353.5                & 2.5                  & 474.9                & 32.4                 & 20.3                  \\
\textbf{Comp.3}      & 158.8                & 226.6                & 47.5                 & 353.1                & 2.6                  & 469.2                & 30.6                 & 19.1                  \\
\textbf{Average}     & 168.0                & 230.8                & 41.0                 & 353.3                & 2.7                  & 473.3                & 34.1                 & 21.4                  \\
\textbf{SD}     & 8.3                  & 6.3                  & 8.2                  & 0.2                  & 0.3                  & 3.6                  & 4.5                  & 3.1                   \\
                     &                      &                      &                      &                      &                      &                      &                      &                       \\
\textbf{Sample}      & \textbf{T(-5\%)}     & \textbf{T1 ($^\circ$C)}     & \textbf{Loss 1 (\%)} & \textbf{T2 ($^\circ$C)}     & \textbf{Loss 2 (\%)} & \textbf{T3 ($^\circ$C)}     & \textbf{Loss 3 (\%)} & \textbf{Residue (\%)} \\
\textbf{PVDF 1}      & 342.9                & 347.1                & 5.6                  & 466.9                & 56.8                 & x                    & x                    & 36.1                  \\
\textbf{PVDF 2}      & 352.9                & 352.8                & 5.5                  & 471.2                & 57.1                 & x                    & x                    & 36.2                  \\
\textbf{PVDF 3}      & 358.9                & 347.0                & 5.2                  & 470.8                & 57.4                 & x                    & x                    & 36.4                  \\
\textbf{Average}     & 351.6                & 349.0                & 5.4                  & 469.6                & 57.1                 & x                    & x                    & 36.2                  \\
\textbf{SD}     & 8.1                  & 3.3                  & 0.2                  & 2.4                  & 0.3                  & x                    & x                    & 0.2                   \\
                     &                      &                      &                      &                      &                      &                      &                      &                       \\
\textbf{Sample}      & \textbf{T(-5\%)}     & \textbf{T1 ($^\circ$C)}     & \textbf{Loss 1 (\%)} & \textbf{T2 ($^\circ$C)}     & \textbf{Loss 2 (\%)} & \textbf{T3 ($^\circ$C)}     & \textbf{Loss 3 (\%)} & \textbf{Residue (\%)} \\
\textbf{RT-28HC}     & 179.8                & 224.5                & 99.0                   & x                    & x                    & x                    & x                    & 1.0                     \\
\multicolumn{1}{l}{} & \multicolumn{1}{l}{} & \multicolumn{1}{l}{} & \multicolumn{1}{l}{} & \multicolumn{1}{l}{} & \multicolumn{1}{l}{} & \multicolumn{1}{l}{} & \multicolumn{1}{l}{} & \multicolumn{1}{l}{}  \\
\textbf{Sample}      & \textbf{T(-5\%)}     & \textbf{T1 ($^\circ$C)}     & \textbf{Loss 1 (\%)} & \textbf{T2 ($^\circ$C)}     & \textbf{Loss 2 (\%)} & \textbf{T3 ($^\circ$C)}     & \textbf{Loss 3 (\%)} & \textbf{Residue (\%)} \\
\textbf{PVDF Powder} & 421.6               & 411.7                & 4.8                & 478.2                    & 62.9                    & x                    & x                    & 32.3                
\end{tabular}

\label{Tab:TGA}
\end{table}

\subsection{PCM Retention Capacity}

Our previous work demonstrated that hierarchical porosity can be used to enhance the anti-leakage performance of shape-stabilized PCMs \cite{grosu2020hierarchical,grosu:2021}. In particular, it was shown that secondary smaller porosity located at the walls of the main large pore significantly enhances capillary forces in the main large pore. It allows achieving a high volume of impregnated PCM, taking advantage of large pores, while avoiding leakage issues taking advantage of the strong capillary force of the smaller pores. This work exploits the benefits of hierarchical porosity by using a microchannel with porous walls (see Figure \ref{SEM}).

\begin{figure}
\centering
\includegraphics[width=0.50\linewidth]{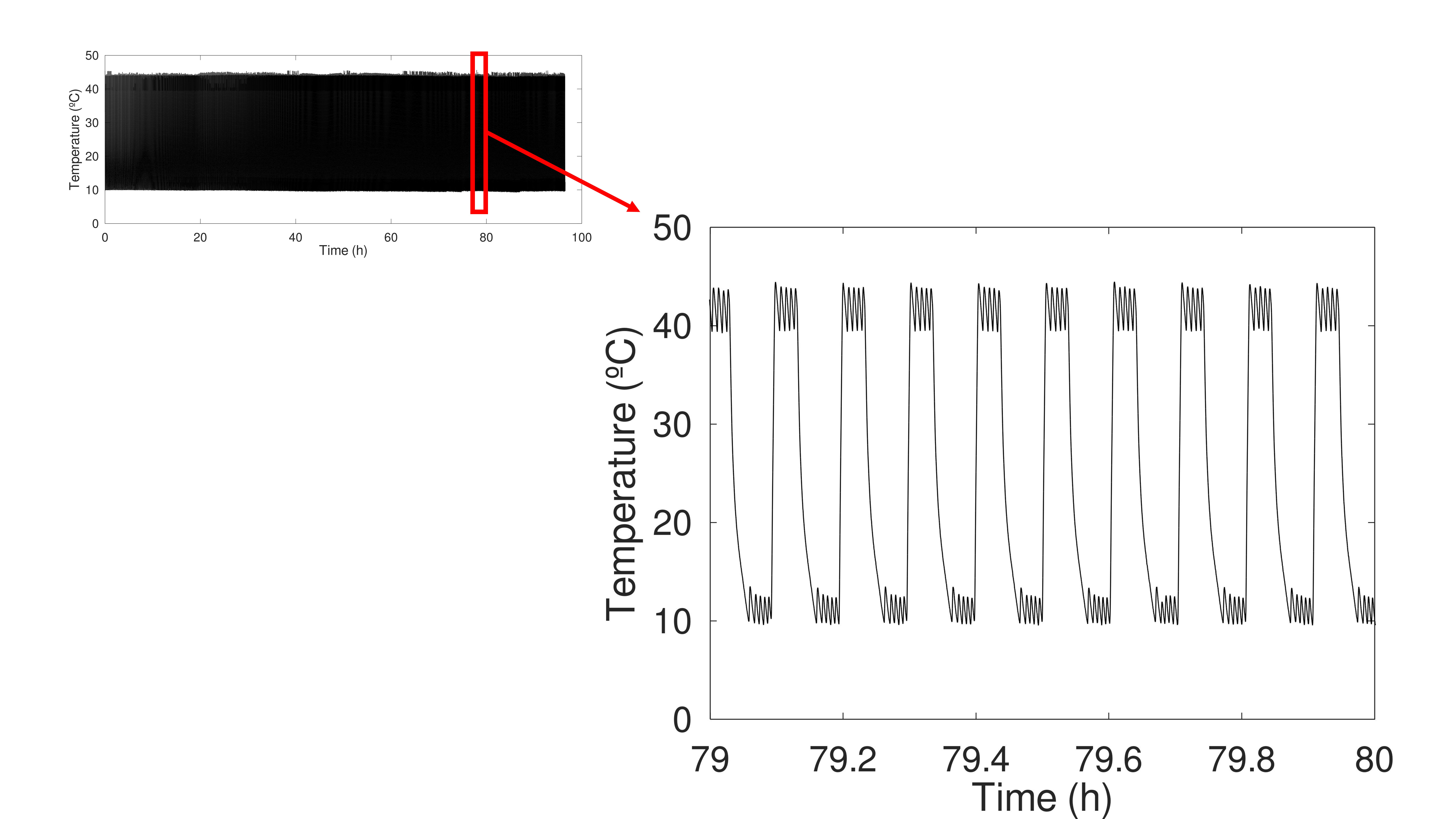}%
\caption{Temperature profile of the thermal cycles used for the leakage test.}
\label{fig:Cycles}
\end{figure}

The produced composite fibre was submitted to 1000 heating/cooling cycles from 10 to 40 $^\circ$C with 120 seconds of isothermal periods at both temperatures. In total, sample was subjected to thermal cycling for almost 100 h. An extract of the temperature profile of the fibre during the cycles is presented in Figure \ref{fig:Cycles}. As it can be seen in Table \ref{Tab:Leakage}, the mass lost by the sample is 0.84\% after the first 50 cycles and stays below 3.5\% after 1000 cycles. Even though no precautions were undertaken to close the ends of the fibers, the results indicate that the composite fibre efficiently retains the molten PCM inside their structure, prevents leakage and maintains thermal storage capacity.

\begin{table}[]
\caption{Mass loss values during the leakage test}
\begin{tabular}{lllll}
Nº Cycles      & 0 & 50   & 100  & 1000 \\
Mass loss (\%) & 0 & 0.84 & 1.06 & 3.47
\end{tabular}

\label{Tab:Leakage}
\end{table}

\subsection{Infrared Analysis of the Thermal Performance} \label{Sec.IR}

\begin{figure}
\centering
\begin{minipage}[h]{0.49\linewidth}
\center{\includegraphics[width=1\linewidth]{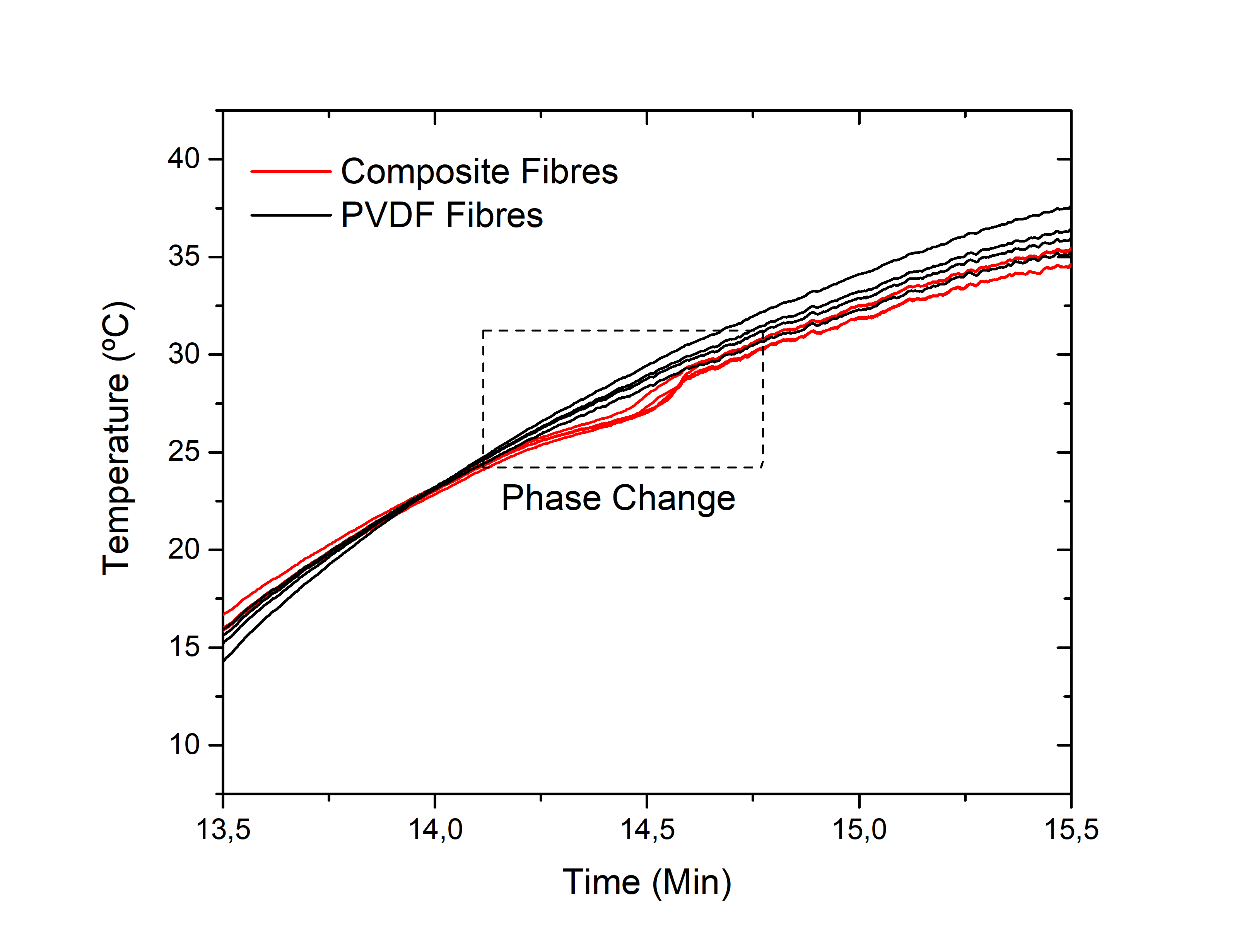} \\ (a)}
\end{minipage}
\begin{minipage}[h]{0.49\linewidth}
\center{\includegraphics[width=1\linewidth]{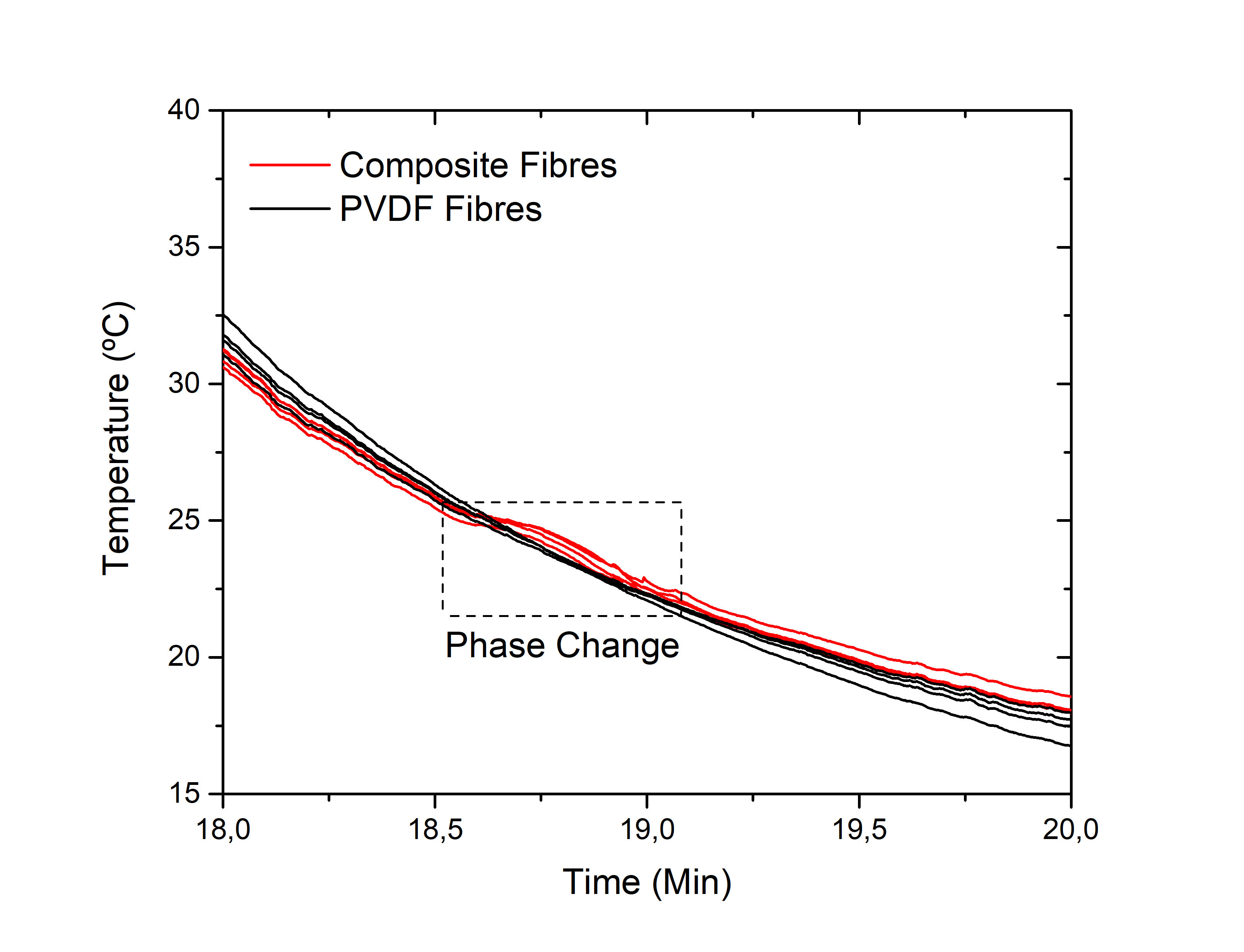} \\ (b)}
\end{minipage}
\caption{Thermal profiles of the fibres during heating (a) and cooling (b).}\label{IRHeating/Cooling}
\end{figure}

\begin{figure}
\centering
\begin{minipage}[h]{0.49\linewidth}
\center{\includegraphics[width=1\linewidth]{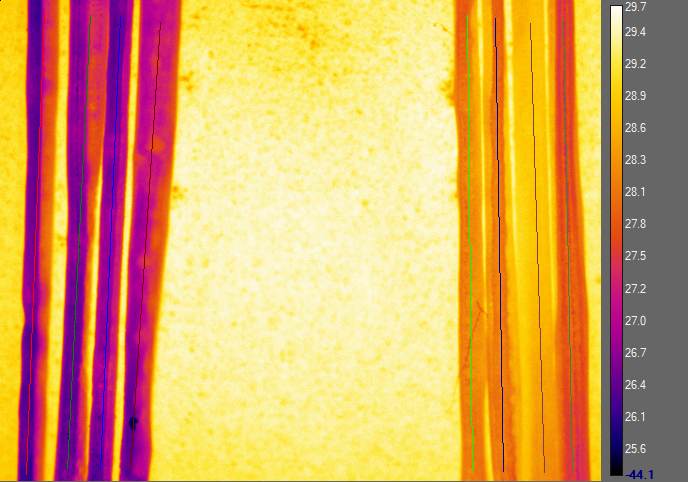} \\ (a)}
\end{minipage}
\begin{minipage}[h]{0.49\linewidth}
\center{\includegraphics[width=1\linewidth]{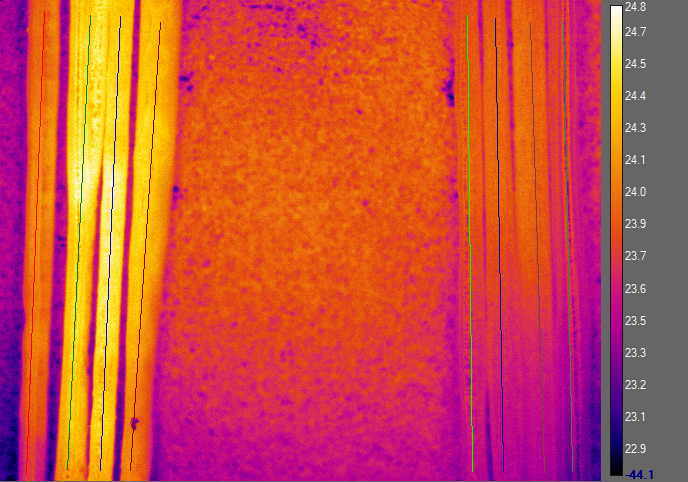} \\ (b)}
\end{minipage}
\caption{Thermal images of the fibres during melting (a) and crystallization (b) processed.}\label{IRHeating/CoolingPhoto}
\end{figure}

As described in section \ref{Thermal performance}, a first data processing was carried out to obtain the temporal evolution of the apparent temperature of each fiber (pure PVDF and composite fibers) during the different cycles of heating/cooling. The results obtained during one cycle of (a) heating and (b) cooling are shown in Figure \ref{IRHeating/Cooling} and \ref{IRHeating/CoolingPhoto}. One can notice that despite an uncertainty on the real emissivity of the fibers, the melting and crystallization temperatures are in agreement with those obtained by standard DSC measurements. 
In addition, the second data processing, based on a sliding subtraction of the recorded images, allows in a simplified way to show localization of the phase change along each fiber (see videos 1 and 2 provided in supplementary material, with and without image processing). Thanks to the large observation scale offered by the IR camera, it could be seen that, during the first cycle, the phase change is highly localized in space, which implies a heterogeneous distribution of the PCM. This observation is also confirmed by the SEM images, DSC and TGA results (see section \ref{Sec.Morphology}, section \ref{Sec.DSC} and section \ref{Sec.TGA}). 
After this first melting/crystallization cycle, the study of the composite fibers by IR thermography provides additional information to that obtained during the previous tests. In fact, from the second heating/cooling cycle, phase change do not exhibit strong localization. That is the result of PCM homogenization and distribution inside each fiber (see videos 3 and 4 provided in supplementary material, with and without image processing). This homogenization is due to the flow of PCM inside the fibers driven by capillary forces when it is in the liquid phase.

\subsection{Mechanical Properties}

\begin{figure}
\centering
\includegraphics[width=0.50\linewidth]{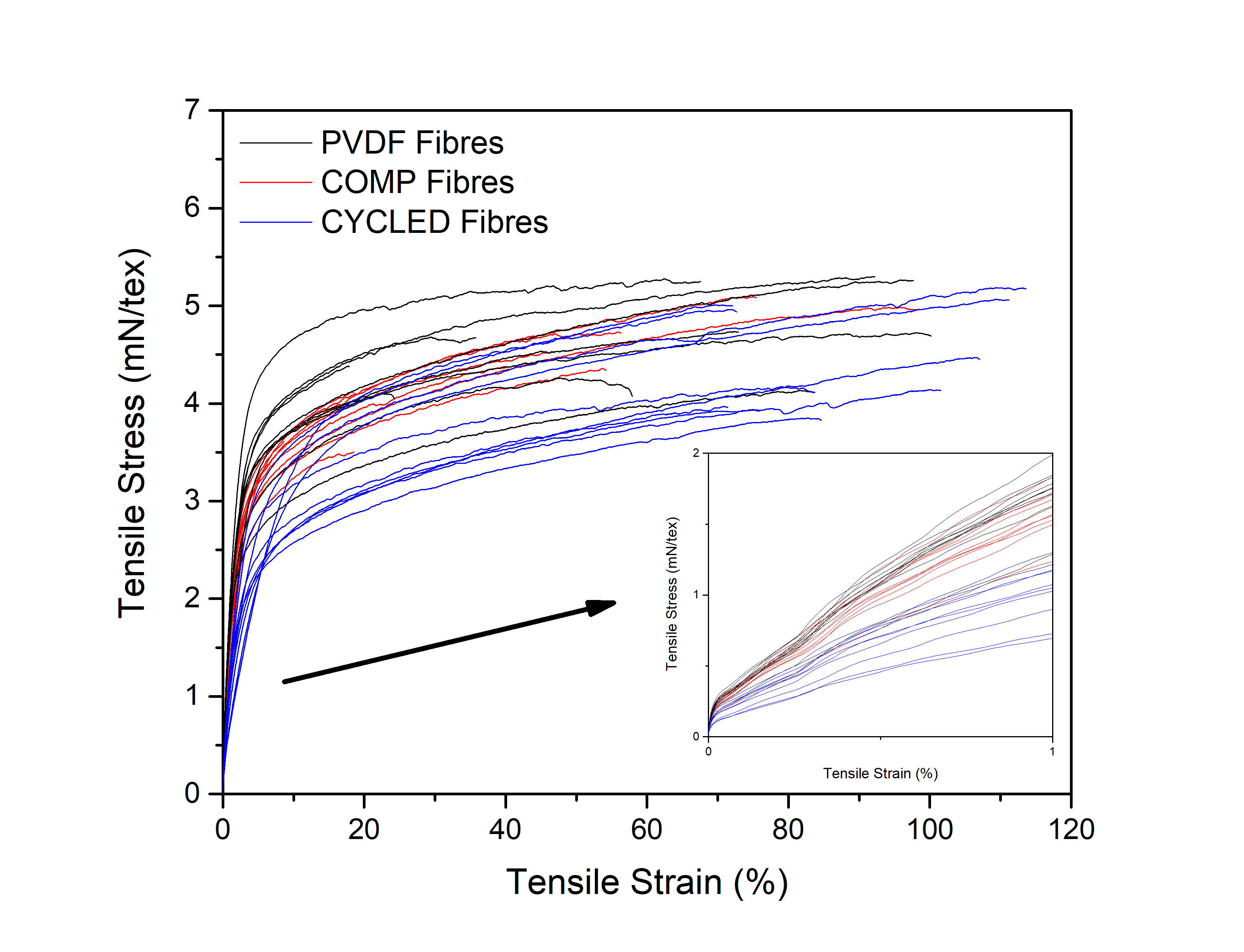}%
\caption{Tensile (Stress VS Strain) profiles of the fibres.}
\label{fig:Tensile}
\end{figure}

The effect of the paraffin content on the mechanical properties of the PVDF fibres is presented in Figure \ref{fig:Tensile} and Table \ref{Tab:TENSILE}, where the mean values for 10 different samples of each fibre are given.
The tensile stress of fibers can be obtained by taking into account the ratio of force corresponding to cross-sectional area of the fibres (Pa) or taking into account the ratio of load to the linear density of the fibres (N/tex). For the case of the fibres produced here, which are hollow and porous, the N/tex units have been selected. This way, errors in the results due to high sensibility to the diameter measurements of the fibres are avoided \cite{Islam2019Tensile}. Each fibre sample was weighed and measured to obtain the linear density of every gauge. 

The stress-strain curves of all the tested PVDF and composite fibres are shown in Figure \ref{fig:Tensile}. Results show that in general, the elastic modulus of as prepared composite fibre is slightly lower than the one of PVDF fibre, decreasing from 1.54 to 1.41 mN/tex. The same happens for the stress and strain values at breakpoint, wich are 4.65 mN/tex and 65\% for pure PVDF fibres while they decrease to 4.18 mN/tex and 42\% for the composite fibres. Nevertheless, these values are comparable to the ones obtained by other authors \cite{zhang2018microfluidic}.
Thus, the presence of paraffin leads to a slight detriment of the mechanical properties of as prepared composite fiber. As it could be seen in the SEM images, the axial structure of the fiber is formed by PCM slugs of different sizes and PVDF walls. The presence and inhomogeneity of the paraffin inside the fibres makes their mechanical properties negatively affected. Nonetheless, the spread of the paraffin inside fibers after 1000 melting/crystallization cycles, visually confirmed by SEM (see Figure ???????) and by IR thermography (see Section \label{Sec.IR}), considerably improve elastically of the composite fibers and strain at breakpoin. The stress at breakpoin for "cycled" fibers is also improved as compared to as prepared composite fibres and is equal to 4.47 mN/tex (see Table \ref{Tab:TENSILE}).

\begin{table}[]
\caption{Elastic modulus and breakpoint tensile properties of the pure PVDF and Composite fibres}
\begin{tabular}{ccccccc}

              & Elastic Modulus & SD       & Stress at Breakpoint & SD       & Strain at breakpoint & SD    \\
              & (mN/tex)        & (mN/tex) & (mN/tex)             & (mN/tex) & (\%)                 & (\%)  \\
PVDF Fibres   & 1,54            & 0,19     & 4,65                 & 0,47     & 65                & 29 \\
COMP Fibres   & 1,41            & 0,15     & 4,18                 & 0,72     & 42                & 29 \\
CYCLED Fibres & 1,00            & 0,21     & 4,47                 & 0,53     & 89                & 17

\end{tabular}

\label{Tab:TENSILE}
\end{table}

\section{Conclusions}
This study presents a new approach for PCM encapsulation into polymeric fibers. The main aim of the study is to provide a simple and highly scalable technology for leakage free PCM production. The aim was achieved by combining core ideas of such methods as emulsion electrospinning, microfluidics and solvent extraction. Based on the obtained results the following conclusions can be drawn:

-The method allows to produce hollow fibers of pure PVDF or composite ones containing paraffin. Paraffin distribution inside the fibers is non-homogeneous and its content varies from 30 to 47.5\% that was measured by DSC and TGA;

-Both higher Cp values and lower peak temperatures of PVDF fibers degradation indicate lower crystallinity of the material after solvent extraction process;

-The produced composite fibers revealed outstanding retention capacity of molten PCM. After 1000 melting/crystallization cycles the mass loose of the PCM did not exceed 3.5\%;

-IR thermography was used to visualize the thermal response of the composite fibers. During the first melting cycle, a strong localization of PCM inside fibers was detected. During the second melting cycle, thermal response was spread over the fibers surface. Consequently, PCM getting spread inside the fibers homogenizing thermal response of composite.

-Measurements of the mechanical properties of the PVDF and composite fibers have shown that both are very flexible. Elongation of the fiber can go up to 100\% before the break up. However, average elastic modulus reduces from 1.54 to 1.41 mN/tex and average stress at breakpoint reduces from 4.65 to 4.18 mN/tex for PVDF and composite fibers respectively. Thus, the presence and  inhomogeneity of the paraffin inside the fibre lead to slightly worse mechanical properties of the composite. However, the spread of the paraffin inside the fibers after thermal cycling improve their mechanical properties.

\section*{Supplementary material}

Video 1: first heating cycle without image processing (\url{https://drive.google.com/file/d/1odyoNPBSOPWTRxQlCqEbaQihq3yzYRGb/view?usp=sharing})

Video 2: first heating cycle with image processing (\url{https://drive.google.com/file/d/1_GUnC1f6skxkV5nkXdAOXg7o-X67LCPK/view?usp=sharing})

Video 3: second heating cycle without image processing (\url{https://drive.google.com/file/d/1N1yTR5DJRxxKNjfQ-XuY4m4gJSy0g_BN/view?usp=sharing})

Video 4: second heating cycle with image processing (\url{https://drive.google.com/file/d/1m-phHOq38iPkMVjTbi_vcSZBUlN_Qwmf/view?usp=sharing})

\section*{CRediT}
Mikel Duran: Writing original draft, Formal analysis, Investigation, Visualization, Writing-review \& editing. Artem Nikulin: Conceptualization, Writing original draft, Methodology, Formal analysis, Supervision, Investigation,
Visualization, Writing-review \& editing. Jean-Luc Dauvergne: Methodology, Formal analysis, Investigation, Visualization, Writing-review \& editing. Angel Serrano: Formal analysis, Writing-review \& editing. Yaroslav Grosu: Formal analysis, Writing-review \& editing. Jalel Labidi: Supervision, Writing-review \& editing. Elena Palomo del Barrio: Supervision, Writing-review \& editing, Funds acquisition.

\section*{Declaration of Competing Interest}
The authors declare that they have no known competing financial interests or personal relationships that could have appeared to influence the work reported in this paper.

\section*{Acknowledgements}

The authors are grateful for the financial support from SWEET-TES project (RTI2018-099557-B-C21), funded by FEDER/Ministerio de Ciencia e Innovación - Agencia Estatal de Investigación and Elkartek CICe2020 project (KK-2020/00078) funded by Basque Government. Mikel Duran Lopez would also like to thank the Department of Education, Linguistic Politics and Culture of the Basque Country government for the granted pre-doctoral contract ({PRE\_2019\_1\_0154}).

\newpage

\afterpage{

\bibliographystyle{elsarticle-num-names} 

\bibliography{references}
}

\end{document}